\documentclass{aa}  

\usepackage{txfonts}
\usepackage{graphicx,lipsum}
\usepackage[percent]{overpic}

\usepackage{amssymb,amsmath}
\usepackage[version=4]{mhchem}
\usepackage{newtxtext}
\usepackage[varvw]{newtxmath}
\usepackage{xcolor}
\usepackage{caption,subcaption}
\usepackage{chemfig} 
\usepackage{hyperref}

\newcommand{\ignore}[1]{}

\def\aap {{A\&A}}

\def\apjl {{ApJL}}

\def\apjs {{ApJS}}

\begin{document}

   \title{Interior redox state effects on the stability of secondary atmospheres and observational manifestations: LP 791-18\,d as a case study for outgassing rocky exoplanets}
  \titlerunning{Interior redox state effects on the stability of secondary atmospheres and observational manifestations}

   \author{Leonardos Gkouvelis\inst{1}\thanks{leo.gkouvelis@physik.lmu.de}, Francisco J. Pozuelos\inst{2}, Thomas Drant\inst{3},Mohammad Farhat\inst{4,5}, Meng Tian\inst{1},Can Ak{\i}n\inst{1}
          }
    \authorrunning{ Gkouvelis et al. }

   \institute{
   $^1$Ludwig Maximilian University, Faculty of Physics, University Observatory,
Scheinerstrasse 1, Munich D-81679, Germany\\
   $^2$Instituto de Astrof\'isica de Andaluc\'ia (IAA-CSIC), Glorieta de la Astronom\'ia s/n, 18008 Granada, Spain\\
   $^3$ETH Zürich, Center for Origin and Prevalence of Life, Department of Earth and Planetary Sciences, Institute for Geochemistry and Petrology, 8092 Zurich, Switzerland\\
   $^4$Department of Astronomy, University of California, Berkeley, Berkeley, CA 94720-3411, USA. \\  $^5$Department of Earth and Planetary Science, University of California, Berkeley, Berkeley, CA 94720-4767, USA.\\
    }

   \date{Received \today}

  \abstract
  { 
Recent advances in space-based and ground-based facilities now allow the atmospheric characterization of a selected sample of rocky exoplanets. These atmospheres offer key insights into planetary formation and evolution, but their interpretation requires models that couple atmospheric processes with both the planetary interior and the surrounding space environment. This work focuses on the Earth-size planet LP 791-18\,d, which is estimated to receive continuous tidal heating due to the orbital configuration of the system; thus, it is expected to exhibit volcanic activity. Using a 1D radiative-convective model coupled with chemical kinetics and an outgassing scheme at the lower boundary, we simulated the planet’s atmospheric composition across a range of oxygen fugacities, surface pressures, and graphite activities. We estimated the mantle temperature of $\approx$1680 - 1880 K, balancing the competing contributions of interior tidal heating and convective cooling. Our results show that the atmospheric mean molecular weight gradient is controlled by oxygen fugacity rather than bulk metallicity.  Furthermore, we used the atmospheric steady-state solutions produced from the interior redox state versus surface pressure parameter space, and explored their atmospheric stability. We find that stability is achieved only in highly oxidized scenarios, $fO_2-IW\gtrsim 2$, while reduced interior states fall into the hydrodynamic escape regime with mass loss rates on the order of $\approx 10^5-10^8$ $\text{kg/s}$. We argue that scenarios with reduced interior states are likely to have exhausted their volatile budget during the planet's lifetime. Furthermore, we predict the atmospheric footprint of the planet's interior based on its oxidation state and assess its detectability using current or forthcoming tools to constrain the internal and atmospheric composition. We show that the degeneracy between bare rock surfaces and thick atmospheres can be resolved by using three photometric bands to construct a color–color diagram that accounts for potential effects from photochemical hazes and clouds. For JWST/MIRI, this discrimination is possible only in the case of highly oxidized atmospheres. The case of LP 791-18\,d enables the investigation of secondary atmosphere formation through outgassing, with implications for similar rocky exoplanets. Our modeling approach connects interior and atmospheric processes, providing a basis for exploring volatile evolution and potential habitability. 
 }


 \maketitle

\section{Introduction}\label{sec:intro}

In the last 30 years, almost 6000 exoplanets have been detected\footnote{According to the NASA Exoplanet Archive, consulted in January 2025.}, and atmospheric characterization has started to become possible for a subset of the known population. Due to observational constraints, to date, it has been possible to measure the atmospheres of gas giants, sub-Neptunes, and a few super-Earths spectroscopically. A few selected candidates of approximately Earth's size have been studied \citep[e.g.,][]{Kreidberg2019}, with the Spitzer InfraRed Array Camera (IRAC) and, more recently, with the James Webb Space Telescope (JWST) \citep[see, e.g.,][]{Zieba2023,Ducrot2024,Xue2024,Bello-Arufe2025}. However, little is still known about the compositions of terrestrial exoplanet atmospheres or even of their existence. This gap can only be addressed through direct atmospheric characterization via transmission spectroscopy, emission spectroscopy, or reflected light observations \citep[see, e.g.,][]{Seidler2024,Hammond2025}, and near-future JWST observations will be dedicated to this effort.

Gas giants retain nebular primary atmospheres, and their composition is directly connected to their formation history, while rocky planets develop secondary atmospheres, which are acquired through various processes, including accretion of nebular gas \citep{Sasaki1990} and devolatilization of impactors \citep{Kuwahara2015}, but mainly through outgassing during their initial magma state and subsequent volcanic degassing \citep[see, e.g.,][]{Abe2011,Carlson2014,Schaefer2017,Gaillard2014}. In this context, we cannot directly connect their atmospheres to the conditions at their formation \citep{Brachmann2025}, at least not without first understanding other intermediate mechanisms.

The geochemical outgassing processes responsible for the birth and evolution of secondary atmospheres occur most intensively in the early stages of planetary evolution, when the interior and surface are hot enough and are often in the magmatic-melt phase. If a planet experiences high tidal energy dissipation, outgassing processes can persist as long as the volatile budget of the interior remains unexhausted, with volcanic activity and volatile cycles playing an important role in the planet's overall evolution over its lifetime.  
Rocky exoplanets that are estimated to exhibit high outgassing activity due to volcanism or those covered entirely or partially by magma oceans offer a rare opportunity to explore the conditions under which planets can develop secondary atmospheres. Rocky exoplanets with high surface and atmospheric temperatures offer enhanced observability in both transmission and emission spectroscopy. In transmission, elevated temperatures can partially offset the suppression of spectral signals caused by high mean molecular weight, increasing the scale height. In emission, higher temperatures enhance the thermal contrast with the host star. These effects are especially pronounced for close-in planets orbiting cool stars, such as M dwarfs, where the small stellar radii amplify transit signals and facilitate the characterization of even tenuous secondary atmospheres dominated by heavy molecules using current instrumentation. Empirical evidence is essential for unraveling the complex interplay between a rocky planet's initial volatile inventory and the atmospheric loss and replenishment processes that shape its evolution. Observations of carefully selected candidates, combined with numerical simulations exploring a broad parameter space, will provide critical test cases for elucidating the observable manifestations of atmospheric processes. Typically, candidates with equilibrium temperatures up to \( 1000 \) K are preferred, as their close proximity to their host stars not only ensures high temperatures but also makes them likely to be in a tidally locked configuration, which may result in significant tidal heating \citep[see, e.g.,][]{Seligman2024}. 

In this context, LP 791-18\,d is an Earth-size planet orbiting the cool M6 dwarf LP 791-18 \citep{Crossfield2019,Peterson2023} and is part of a coplanar system. This configuration presents a unique opportunity to study an exo-Earth alongside a sub-Neptune that may have retained its gaseous or volatile envelope. Given its potential for atmospheric characterization with JWST, this planet was selected as a case study for this work. Furthermore, having a sub-Neptune in the same system makes it possible to conduct comparative planetology. The gravitational interaction with the sub-Neptune prevents the complete circularization of LP 791-18\,d’s orbit, resulting in continued tidal heating of the interior, possibly inducing high volcanic activity. This continuous tidal heating must be balanced by convective cooling through the surface, which can lead to significant outgassing activity. The estimation of the mantle equilibrium temperature from \cite{Peterson2023} for different melt-to-rock fractions ranges from \( 1630 \) to \( 1670 \) K. In the next section, we demonstrate that this calculation likely underestimates the magma temperature and that the interior thermal equilibrium is expected to reach higher temperatures by a few hundred kelvins. Observations with the JWST during Cycle 3 have been approved for LP 791-18\,d (\citet{Benneke2024}). The aim of this observation is to achieve a \( 4\sigma \) confidence level in distinguishing between a bare-rock scenario and a \( \text{CO}_2 \)-rich atmosphere capable of efficient heat redistribution, even under conditions that might include Venus-like clouds. Observations are planned for five visits with the F1500W filter and the SUB256 sub-array, a \( 3\,\mathrm{\mu m} \) band-pass centered at \( 15\,\mathrm{\mu m} \), which covers a strong absorption feature from \( \text{CO}_2 \). The detection of an atmosphere on LP 791-18\,d would provide the first evidence that a temperate rocky planet orbiting an M dwarf can sustain one. All of the above make LP 791-18\,d a benchmark candidate for the exploration of outgassing processes and the evolution of secondary atmospheres on terrestrial planets orbiting M dwarfs. \\  

\section{Approach} \label{sec:approach}

To assess the role of interior redox state on the atmospheric stability and observability of LP~791-18~d, we adopted a modeling framework that connects geochemical outgassing and atmospheric processes, and finally to emission spectra and observational diagnostics. In this section, we present a road-map of our approach and outline the steps followed in our theoretical framework.

A key reason for selecting LP~791-18~d as our case study is the ability to estimate its internal thermal state by balancing the energy flux with convective cooling. This calculation is presented in Section~\ref{sec:melt}. Once the internal temperature is determined, it is assumed to correspond to the melt temperature, be it the temperature of surface lavas or the temperature of subsurface magma chambers. This temperature is used to minimize the multi-dimensional parameter space of planetary unknowns, namely the chemical inventory (parameterized by interior oxygen fugacity), surface pressure, melt temperature, and graphite activity. The outgassing rate is parametrized as described in Section~\ref{sec:simulations}. The volatile content of the planet is partitioned between the melt and the atmosphere through equilibration, achieved by minimizing the system’s Gibbs energy. Once the gases are released into the atmosphere, we apply full chemical kinetics, photochemistry coupled with radiative convective model, until a steady-state composition is reached.
In Section~\ref{sec:simulations}, we describe the numerical tools used to perform simulations across a parameter grid defined by our main variables: oxygen fugacity, surface pressure, and graphite activity. Because this parameter space spans a wide range of atmospheric compositions—from highly reducing to highly oxidizing—and from tenuous to massive atmospheres, we aim for a self-consistent approach. For example, we used an analytical approximation for the heat redistribution efficiency and applied mixing length theory to estimate eddy diffusion coefficients. A sensitivity analysis on surface albedo  is also performed to test the robustness and broader applicability of our modeling framework under a broad range of surface properties.

\begin{figure}[ht]
    \centering
    \includegraphics[scale=0.28]{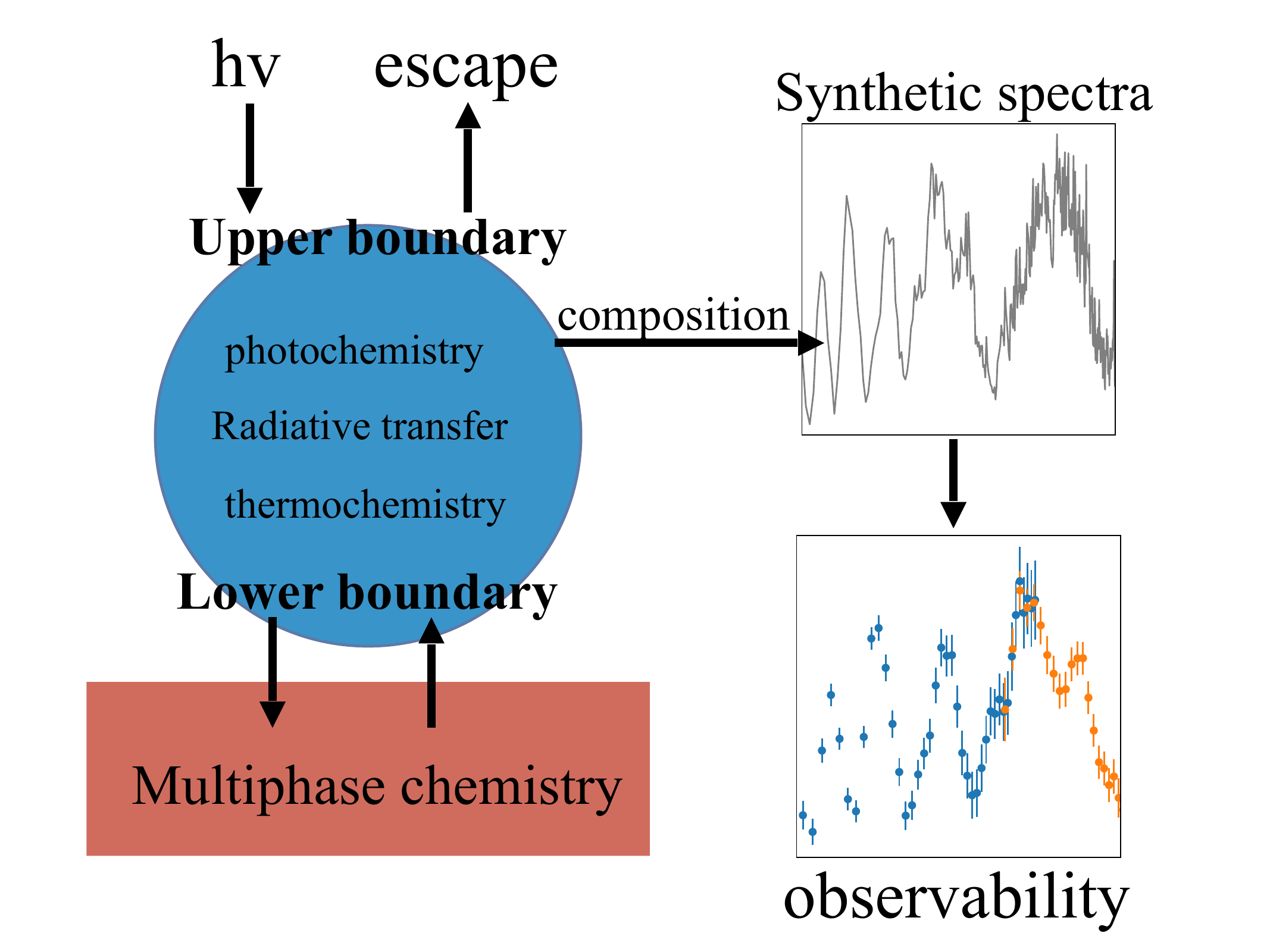}
    \caption{Mind map describing the different parts of our methodology. The simulations are composed of three core pieces coupled together. Radiative convective model, chemical kinetics, and an equilibrium inter-phase multi-phase chemistry between the surface and the atmosphere. Post-numerical simulation analysis is done to produce transmission and emission mock spectra.}
    \label{fig:mind_map}
\end{figure}

Once steady-state atmospheric compositions are obtained, we evaluated their survivability under the background stellar environment in Section~\ref{sec:stability} by estimating its escape regime and  atmospheric mass loss rates. 
The next step in our analysis is the generation of mock spectra to evaluate observational detectability with JWST at selected photometric bands. We produced two categories of mock spectra. The first includes airless scenarios, representing LP~791-18~d after complete atmospheric escape. For this, we used Hapke reflectance and emittance theory, as described in Section~\ref{subsec:bare_rock}. The second category includes the thermal emission spectra of atmospheric scenarios, following the methods in Section~\ref{sec:MOCK}. We then analyzed all spectra by integrating them over JWST bandpasses and discuss the potential for distinguishing atmospheric scenarios from bare rocks.
In addition, we performed parametric radiative transfer simulations that include the effects of photochemical hazes and clouds to assess their impact on our spectral predictions (Section~\ref{sec:hazes}).  Figure~\ref{fig:mind_map} presents a concise mind map illustrating the interplay between the various physical mechanisms incorporated into our simulations and the resulting observable quantities.

In Section~\ref{sec:discussion}, we discuss the implications of our findings for upcoming JWST photometric observations of LP~791-18~d, and infer the possible current state of the planet. We conclude by outlining broader implications for the population of small, rocky exoplanets and suggest promising targets for future studies.

\section{The outgassing nature of LP 791-18\,d and its internal thermal state.} \label{sec:melt}

The possible thermal state of the planet's interior was investigated in \citet{Peterson2023} by balancing the competing contributions of interior tidal heating and convective cooling. Their calculations were performed using the publicly available melt module\footnote{\url{https://github.com/cpiaulet/melt}}  (\citet{Barr2018,Dobos2015,Moore2003}).

\begin{figure}[ht!]
    \centering
    \includegraphics[width=0.5\textwidth,, trim=0cm 0cm 0cm 0cm, clip]{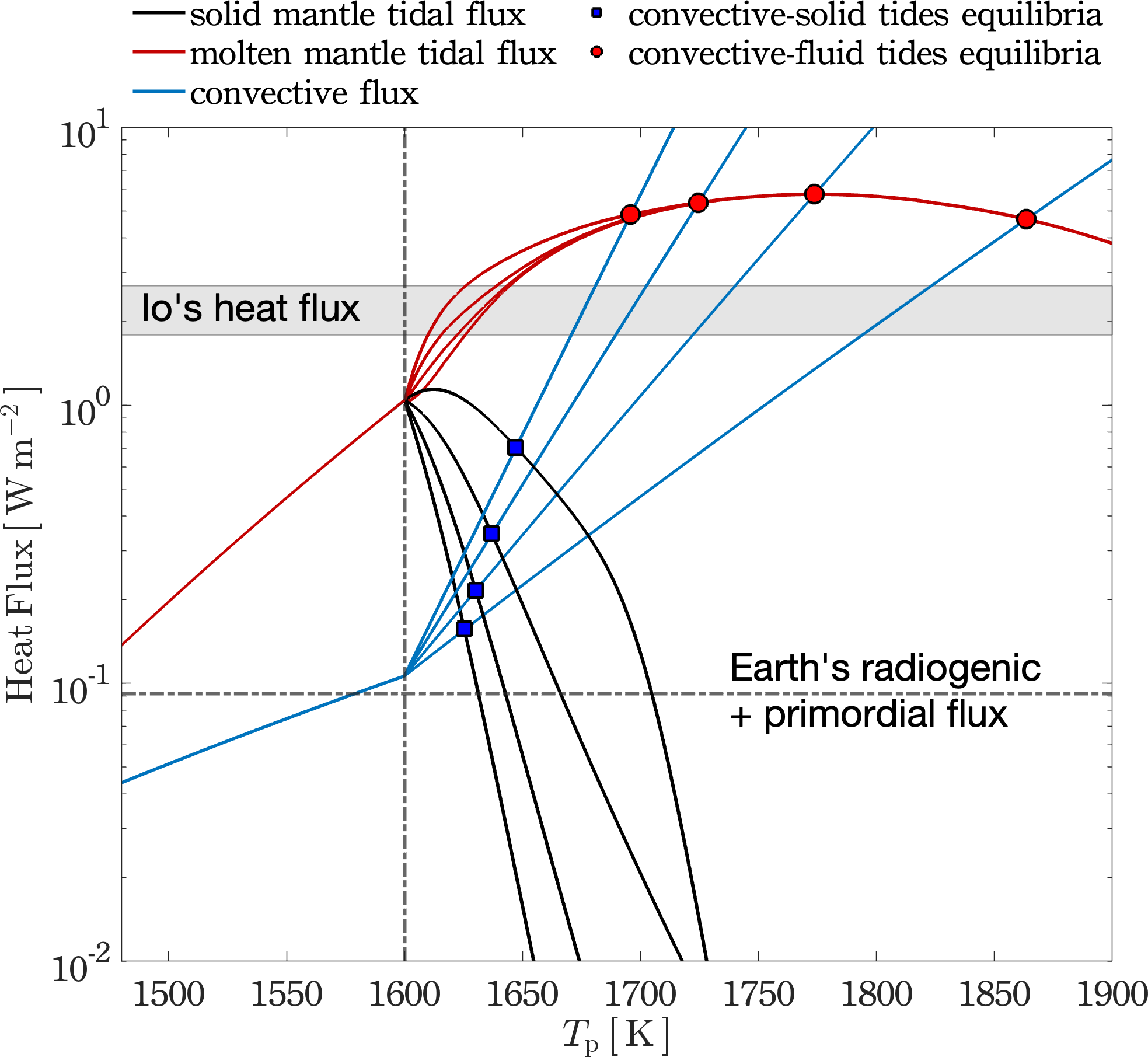}
    \caption{Modeled internal energy balance for LP 791-18\,d. Plotted are convective (blue curves) and tidal (black and red curves) heat fluxes as a function of the mantle's potential temperature. The isothermal mantle solidus is set at $T_{\rm p}=1600{~\rm K}$. We follow the prescription of \citet{Peterson2023} and use the same parameters therein to compute: \textit{i}) the convective flux assuming soft turbulence in the mantle \citep[e.g.,][]{SOLOMATOV201581}, whereby the four sets of curves cover a range of melt fraction coefficients ($B=10,20,30,{\, \rm and}\,40$); and $\textit{ii})$ the tidal heating flux of a Maxwell viscoelastic solid mantle. In addition, we compute the tidal heat flux allowing for the fluid contribution of melt in the mantle, following \citet{farhat2024lava}, using a conservative estimate of the dissipative timescale, $\sigma_{\rm R}=10^{-3}$~s$^{-1}$. Interior thermal equilibria, corresponding to the balance between tidal heating and convective cooling, are shown by the colored markers.  }
    \label{fig:tidal_heating}
\end{figure}

Their results indicated the existence of stable thermal equilibria at mantle temperatures ${\sim1620-1650}\,{\rm K}$, i.e., above the presumed solidus temperature of the mantle. Consequently, the planet is expected to harbor a partially molten mantle that evades complete solidification by eccentricity-driven tidal heating balancing the escaping convective flux. We reproduce this family of interior thermal equilibria obtained by \citet{Peterson2023} in our Figure \ref{fig:tidal_heating}, represented by the blue squares. For these equilibria, the surface flux due to tidal heating is ${\sim0.2-0.8}\,{\rm W\,m^{-2}}$, that is, one order of magnitude greater than the heat flux of the Earth \citep[see, e.g.,][which is mainly attributed to its radiogenic activity]{Davies2010}, and one order of magnitude less than that of Io \citep[see, e.g.,][which is primarily driven by the tidal action of Jupiter's gravitational field]{lainey2009strong}.

These results are based on the assumption that the mantle, at all temperatures, responds to tidal forces as a viscoelastic solid \citep[the specific rheology adopted in \cite{peterson2023temperate} to model viscoelasticity is that of a Maxwell solid; see][for a recent review on solid tidal rheolgies]{bagheri2022tidal}. However, for mantle temperatures beyond the solidus, the increasing melt fraction causes the molten layer to exhibit fluid-like behavior under tidal stresses rather than as a solid. The tidal signature of this rheological transition in molten layers was recently explored by \cite{farhat2024lava} by coupling the equations describing dynamical tides in fluid layers with those describing flexure in a viscoelastic solid. The resulting difference in the tidally generated heat flux between the two models, shown in Figure \ref{fig:tidal_heating}, is summarized as follows. In the framework of a solid mantle, the tidal heat flux is controlled by the mantle’s temperature-dependent viscosity: as the temperature increases, more melt is generated, the mantle's viscosity decreases, and so does the generated heat. This explains the decaying solid tidal flux (black curves in Figure \ref{fig:tidal_heating}) as a function of temperature. In contrast, if the contribution of the fluid within the molten layer is taken into account, at temperatures beyond the solidus, the tidal response increases with increased temperature by virtue of the increased melt fraction. 

Consequently, accounting for fluid tidal heating, interior thermal equilibria of LP 791-18\,d would shift towards higher temperatures (${\sim1680-1880}\,{\rm K})$, depending on the profile of convective cooling. The surface heat flux also increases to ${\sim4.5-6}\,{\rm W\,m^{-2}}$, exceeding that which generates the vigorous volcanic activity on Io \citep[${2.24\pm0.45}\,{\rm W\,m^{-2}}$,][]{lainey2009strong}. Notably, these new equilibria, similar to those in the solid framework, are dependent on the actual mantle phase diagram, specifically on the radial profiles of the solidus and the critical melt fraction at which the rheological transition occurs. In the evident absence of such constraints, one can confidently argue that the equilibrium temperatures obtained in the solid framework can only serve as lower limits to the thermal state of the partially molten mantle, while the enhanced effect of fluid tides can significantly increase the equilibrium temperatures and the resulting surface flux of the planet. This is of utmost significance for atmospheric dynamics since the outgassed pressure scales exponentially with the melt's temperature. 

\section{Numerical modeling}\label{sec:simulations}

This work implements a self-consistent and systematic modeling framework for identifying spectra that can provide insight into the formation of LP 791-18\,d's atmosphere and serve as a proxy for rocky exoplanets in general. Unlike other state-of-the-art models of rocky exoplanet atmospheres to date, we model an outgassing-like planet as an integrated radiation-atmosphere-melt physical system. 

The atmospheric environment in an actively outgassing planetary state is expected to interact with the melt near the surface. The atmospheric composition is determined by the volatile components released through outgassing from the magma's interior. At this hot lower boundary, multiphase chemical reactions are fast enough to assume equilibrium between the first layers of the atmosphere and the melt \citep[see, e.g.,][and the references therein]{Tian2024}. It is important to note that, in the volcanic degassing scenario, equating the atmospheric surface pressure to the outgassing pressure implicitly assumes that gas–melt equilibration always occurs at the surface, rather than deeper within the magma chamber. Once this hot gas is released into the cooler environment of the upper atmospheric layers, it can no longer be considered in equilibrium, and a proper chemical-kinetics treatment is required. We solved the chemical kinetics continuity equation as prescribed by \cite{Hu2012} and \cite{Tsai2017}, including photochemistry in the upper layers. Hence, the chemical kinetics equation with eddy mixing, molecular diffusion, and production and loss terms for each species is mathematically described as:
\begin{equation}
    \frac{\partial C_i}{\partial t} =\Phi_i + P_i - L_i
\end{equation}
where the left-hand side of the equation represents the rate of change in the concentration of species \( i \) over time. The $\Phi_i$ on the right-hand side represents the vertical transport of species, which is given in more detail below. Finally, \( P_i \) and \( L_i \) are the chemical production and loss mechanisms of each species per altitude grid. In this work, we utilized the VULCAN model \citep[see,][]{Tsai2017,Tsai2021} to self-consistently simulate atmospheric chemical kinetics, vertical mixing (including both molecular diffusion and eddy transport), and photochemistry. Specifically, VULCAN solves the full chemical kinetics continuity equation, accounting explicitly for diffusive vertical transport processes alongside chemical production and loss terms for each atmospheric species. VULCAN includes a database of over 700 neutral reactions. The photochemical calculations within VULCAN also include a built-in UV radiative transfer module with multiple scattering to simulate photodissociation processes accurately. Due to its versatility, this powerful model can simulate any solar or extrasolar atmosphere with relatively minor customization. Thermochemistry within the atmosphere strongly depends on temperature, which, in turn, is influenced by background radiation. To address this problem, we coupled radiative transfer with chemical kinetics. The governing equation can be written as:
\begin{equation}
     \frac{dI_{\nu}}{ds} = -\alpha_{\nu} I_{\nu} + \eta_{\nu} + \epsilon_{\nu} - \rho_{\nu} I_{\nu} 
\end{equation}
where \( \frac{dI_{\nu}}{ds} \) represents the rate of change of intensity with penetration \( s \) within the atmosphere, \( I_{\nu} \) is the radiation intensity at frequency \( \nu \), \( \alpha_{\nu} \) is the absorption coefficient, \( \eta_{\nu} \) is the emission coefficient, \( \epsilon_{\nu} \) is the scattering coefficient, and \( \rho_{\nu} \) is the reflection coefficient.

The radiative transfer equation is solved with the Helios\footnote{\url{https://github.com/exoclime/HELIOS}} self-consistent radiative transfer code described in \cite{Malik2017} and \cite{Malik2019}. Opacities for atmospheric species were taken from the Data \& Analysis Center for Exoplanets (DACE)\footnote{\url{https://dace.unige.ch/opacityDatabase/}} database. The interaction between different computational schemes was established to study realistic couplings between interior and atmospheric environments. The coupling of the numerical codes is described in \citet{Drant2025}. \\  

\subsection{Vertical transport} 

Vertical transport and mixing of atmospheric species is an important component of our framework. The balance between eddy mixing and molecular diffusion determines the location of the homopause layer. This, in turn, defines the spatial region where molecular diffusion acts such that it can modify the mean molecular weight at the base of the possible planetary wind, $R_{xuv}$  (see section \ref{sec:stability} for details).

For the eddy diffusion coefficient, $K_{zz}$, we wanted to employ a self-consistent approximation that will be flexible across the wide range of atmospheric compositions simulated in this work.  We based our approach on the Mixing Length Theory (MLT), assuming convection dominates vertical transport below the radiative region. For each layer, we compute eddy diffusion as: 
\begin{equation}
    K_{zz} = \ell^2 N,
\end{equation}
where $\ell$ is the vertical mixing length and $N$ is the Brunt–Väisälä frequency. The mixing length is assumed to be a fraction of the atmospheric scale height
\begin{equation}
    \ell = \alpha H, \quad \text{with} \quad H = \frac{k_B T}{\mu m_H g},
\end{equation}
where $\alpha$ is the mixing length parameter (in this work we assume $\alpha = 0.1$ across all of our simulations, which was chosen by benchmarking with the background typical atmospheres for Venus, Earth and Mars), $c_p$ is the local specific heat capacity at constant pressure for the atmospheric composition, $T$ is the local temperature, and $g$ is the gravitational acceleration. The Brunt–Väisälä frequency:
\begin{equation}
    N = \sqrt{ \frac{g}{T} \left( \frac{dT}{dz} + \frac{g}{c_p} \right) }.
\end{equation}
 The specific heat capacity and mean molecular weight $\bar{\mu}$ are determined from the local atmospheric composition.  This yields a self-consistent $K_{zz}$ profile varying with altitude, temperature, and atmospheric composition. 
 
For the molecular diffusion, we followed the methodology below. Since we solve self-consistently the photochemistry from the outgassing species from the lower boundary, we are agnostic to an a priori background dominant atmospheric gas around homopause; for example, \citet{Tsai2021} includes molecular diffusion of species to an a priori selection of background gas. In this work, we have to follow a full multi-component treatment and not assume any dominant background species. This is done by constructing a matrix of binary diffusion coefficients $D_{ij}$ between all pairs of species as a function of altitude using the Chapman–Enskog formula, corrected by a species-specific $\Gamma_{ij}$ factor (\citet{Chapman1970}): 
\begin{equation}
D_{ij} = \frac{k_B T^{3/2}}{P \, \sigma_{ij}^2 \, \Omega_{ij}},
\end{equation}
where $k_B$ is Boltzmann’s constant, $T$ is the temperature, $P$ is the pressure, $\sigma_{ij}$ is the average collision diameter of species $i$ and $j$, and $\Omega_{ij}$ is the corresponding collision integral. A correction factor $\Gamma_{ij}$, empirically determined or set to 1.0 by default, accounts for non-ideal interactions (\citet{Banks1973}).
The binary diffusion coefficients are then used to compute species-specific diffusion coefficients $D_i$ using Wilke’s Rule
\begin{equation}
D_i = \left( \sum_{j \neq i} \frac{X_j}{D_{ij}} \right)^{-1},
\end{equation}
where $X_j$ is the mole fraction of species $j$. The effective molecular diffusion coefficient $D_{\rm eff}$ of the mixture is then calculated as a weighted average
\begin{equation}
D_{\rm eff} = \sum_i \phi_i D_i,
\end{equation}
where $\phi_i$ is the mixing ratio of species $i$. The collision diameters $\sigma$ and collision integrals $\Omega$ are taken from standard tabulated values for common molecules (\citet{Carey1988}).

Finally, combining both coefficients in the vertical transport, we can introduce them in equation \ref{eq:continuity} as

\begin{equation}
\Phi_i = -C \left( K_{zz} \frac{\partial f_i}{\partial z} + D_{eff} \left[ \frac{\partial f_i}{\partial z} + f_i \left( \frac{1}{H_i} - \frac{1}{H} \right) \right] \right),
\end{equation}

with the \( \Phi_i \) in \(\mathrm{cm}^{-2}\,\mathrm{s}^{-1}\), \( f_i = \frac{C_i}{C} \) is the mixing ratio of species \( i \), and \( H_i = \frac{k_B T}{m_i g} \) is the scale height of species \( i \).
 
\subsection{Atmospheric heat redistribution}

 An important question in modeling exoplanet atmospheres with 1D models is how to represent the physical processes that govern atmospheric heat redistribution between the day and night hemispheres. A common approach is to modify the planet’s energy budget through a redistribution factor \( f \), such that the dayside brightness temperature \( T_{\mathrm{day}} \) is given by \citep{Burrows2014}:
\begin{equation}
T_{\mathrm{day}}^4 = \left( \frac{R_\star}{d} \right)^2 (1 - \alpha_B) f \, T_\star^4,
\end{equation}
where \( R_\star \) and \( T_\star \) are the stellar radius and temperature, \( d \) is the orbital distance, and \( \alpha_B \) is the Bond albedo. The factor \( f \) encapsulates the efficiency of day–night heat redistribution, ranging from \( f = \tfrac{2}{3} \) for no redistribution (dayside-only re-radiation) to \( f = \tfrac{1}{4} \) for full, uniform redistribution.

In our framework, we explore a wide range of atmospheric compositions with mean molecular weights varying from approximately 2 to 30 amu. To remain physically consistent across this range, we approximate \( f \) using the analytic scaling derived by \citet{Koll2022}, which has been benchmarked against 3D general circulation models. This scaling combines weak-temperature-gradient theory and heat engine arguments to estimate the circulation strength and, consequently, the heat redistribution efficiency. The expression we adopt for \( f \) corresponds to Equation (10) in \citet{Koll2022}, and it smoothly interpolates between the thin- and thick-atmosphere regimes. It thus provides a flexible and physically motivated approximation of the redistribution factor across a wide variety of planetary scenarios.

\subsection{Degassing} 
The above governing equations describe the atmosphere. At the lower boundary, at the surface-atmosphere interface, we bring the system into thermochemical equilibrium with an outgassing melt. The model of \cite{Tian2024} is employed here to calculate outgassing at the lower boundary, taking oxygen fugacity, surface pressure, and graphite activity as inputs, and producing volume mixing ratios of outgassed species (e.g., CO, H$_2$O) as outputs. 

\begin{figure}[ht!]
    \centering
    \includegraphics[scale=0.55, trim=3.85cm 0cm 5cm 0cm, clip]{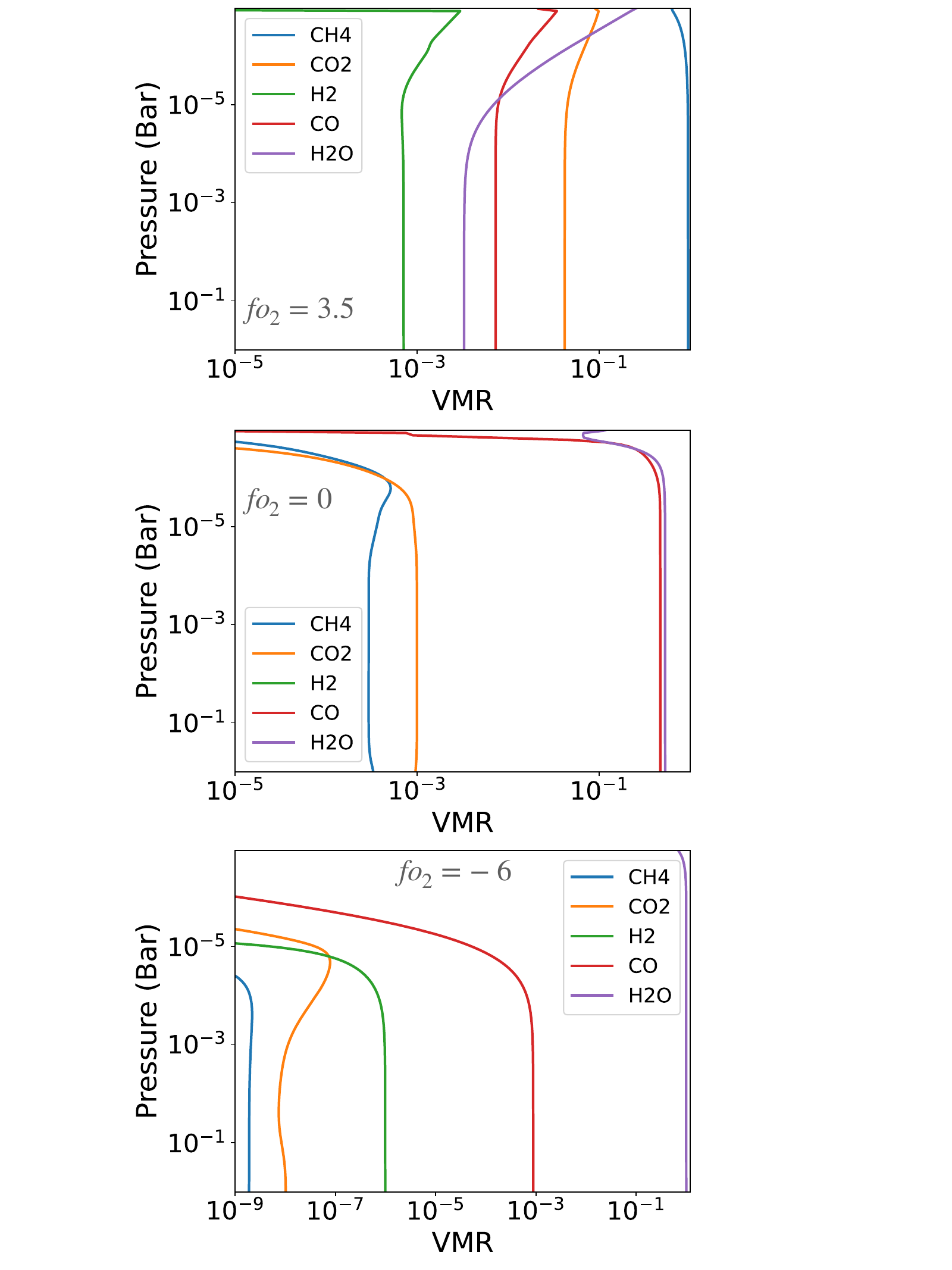}
    \caption{Converged simulations for a surface pressure of $1\text{ bar}$. Volume mixing ratios (VMR) of important greenhouse gases that significantly contribute to emission and transmission spectroscopy are shown. Three representative oxygen fugacity cases are presented to illustrate the transition from a reduced to an oxidized atmosphere.  }
    \label{fig:VMR}
\end{figure}

\begin{figure}[ht!]
    \centering
    \includegraphics[scale=0.255, trim=0.19cm 2.9cm 0 2cm, clip]{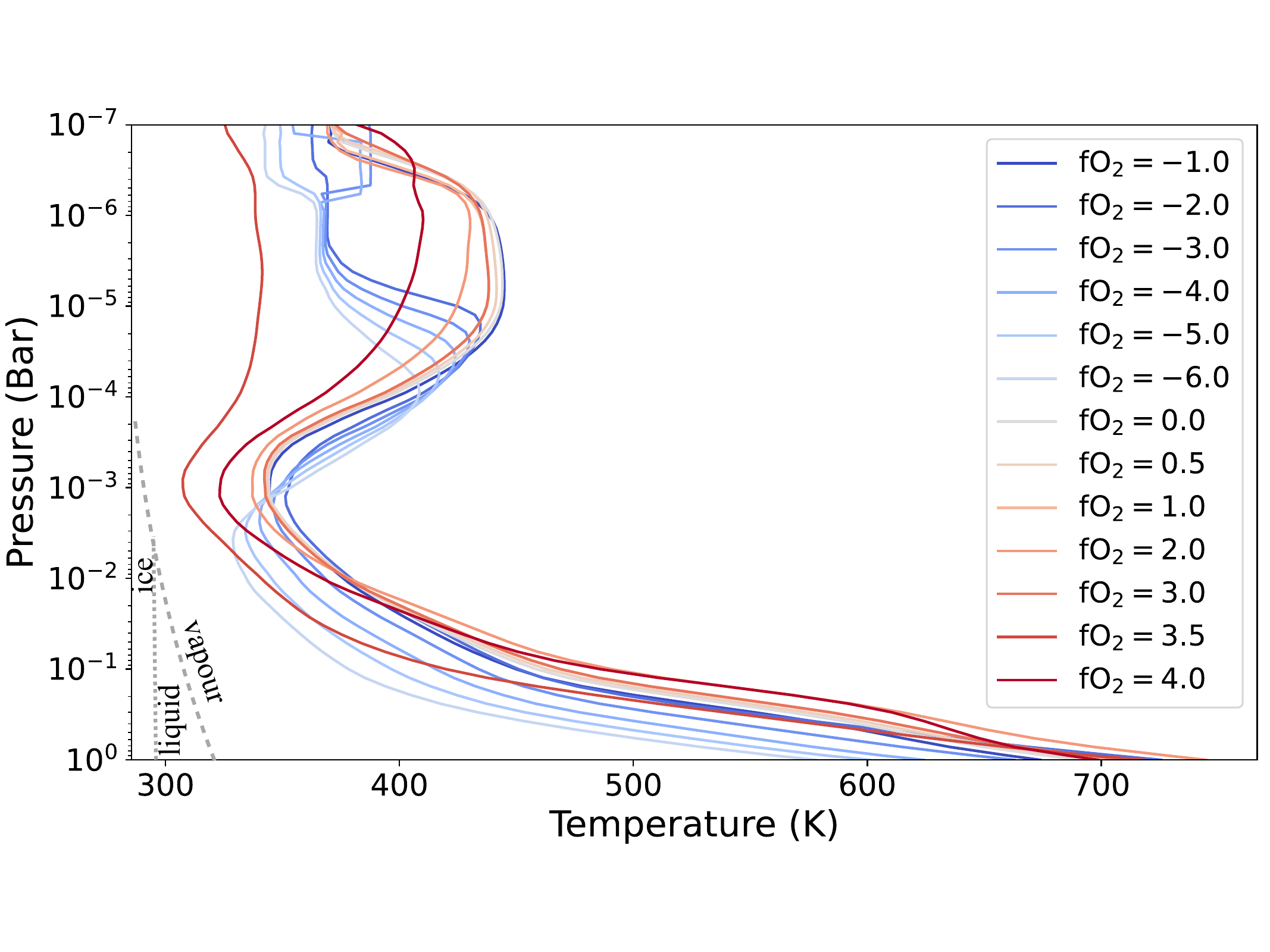}
    \caption{Converged simulations for a surface pressure of $1\,\mathrm{bar}$. The thermal structure is shown for all simulated scenarios within the $1\,\mathrm{bar}$ surface pressure grid. We note the thermal inversion in the upper layers of the simulated atmosphere, caused by infrared absorption due to high water vapor abundances. }
    \label{fig:TP}
\end{figure}

As initial conditions, we prescribed oxygen fugacity, surface pressure, and graphite activity. The graphite activity (\( a_{\rm C} \)) is a hypothetical parameter dictating the carbon content in the outgassing melt, and thus it indirectly affects the carbon content of the outgassed atmosphere. \( a_{\rm C} \) values ranging from \( 10^{-7} \) to \( 10^{-3} \) are explored in this study, with the lower bound corresponding to carbon-poor outgassing and the upper bound representing relatively carbon-rich outgassing \cite{Tian2024}. The temperature is treated as constant for all simulations, with a fixed value of \( 1720 \) K, as explained in Section \ref{sec:melt}. The atmosphere–interior boundary condition described above sets only the starting chemical composition at the atmosphere's lower boundary, assuming local thermochemical equilibrium at the gas–melt interface. It does not, however, directly impose global-scale equilibrium throughout the entire atmospheric column. The vertical atmospheric composition and thermal profiles are instead computed dynamically, driven by chemical kinetics, photochemistry, molecular diffusion, vertical mixing (eddy transport), and radiative–convective processes. Thus, although our lower-boundary conditions implicitly assume equilibrium gas–melt interactions (applicable both to global magma-ocean scenarios and localized volcanic degassing events), the resulting atmospheric structure naturally emerges from these self-consistent atmospheric processes.  We assume the equilibrium occurs either at the surface in active volcanic vents or near the surface inside magma chambers just before degassing. Due to the high melt temperature assumed (1720 K), local chemical equilibrium is a reasonable assumption, and it does not imply a global-scale interface between melt and atmosphere; instead it merely means gas and melt are in equilibrium before the gas escapes melt to contribute to atmosphere growth (e.g. \citet{Bower2022},\citet{Shorttle2024},\citet{Tian2024}).

A constant surface gas emission flux is prescribed in each simulation until convergence to a steady-state atmosphere is reached. This lower boundary source term is included in the chemical continuity equation as a fixed influx of species, scaled by their mixing ratios in the outgassed composition. Our approach follows the simplified parameterization described by \citet{James2018}, in which the total surface outgassing flux $f_j$ of a species $j$ is expressed as:
\begin{equation}
f_j = \dot{V} \cdot \chi_j,
\end{equation}
where $\dot{V}$ is the global magma production rate (in m$^3$\,s$^{-1}$), and $\chi_j$ is the mole fraction of species $j$ released per unit volume of erupted magma. This expression is used to fix the lower boundary condition for the chemical continuity equations and is implemented in our framework as described in \citet{Tsai2017}. Although $\dot{V}$ is treated as constant during the simulations, it can be physically motivated by planetary parameters. The internal energy release from tidal heating provides an estimate of the total energy available to drive mantle melting. Assuming that a fraction $\eta$ of this heat contributes to melt production, and using the latent heat of fusion $L$ and the mantle density $\rho$, the magma production rate can be estimated as:
\begin{equation}
\dot{V} = \frac{\eta \, \dot{Q}}{\rho \, L}
\end{equation}

For LP~791-18\,d,  $\dot{Q}$  and related parameters are taken from tidal dissipation models  provided in Section~\ref{sec:melt}. We estimate a magma production rate of $\dot{V} \approx 20 - 27$ km$^3$ yr$^{-1}$. This value is comparable to the volcanic production rates observed on Earth and Venus, and lies within the plausible range considered in exoplanetary volcanic outgassing studies \citep[e.g.,][]{ Gaillard2014, James2018}.

Once equilibrium is established between the melt and the atmosphere, numerical simulations are executed until a steady state is reached, accounting for thermochemistry and photochemistry. The stellar input flux was taken from the MUSCLES\footnote{\url{https://archive.stsci.edu/prepds/muscles/}} catalog. MUSCLES is a spectral survey of M and K exoplanet host stars covering wavelengths from X-ray and ultraviolet to infrared (\cite{Behr2023}, \citet{Wilson2025}). 

Each simulation reached statistical steady-state conditions after the change in parameters reached a threshold, where the minimum limit for convergence was set to \( \Delta U \approx 10^{-6} \), which was computed at each time step as:

\begin{equation}
\Delta U =\sqrt{ \frac{1 }{ N_{var}}   \displaystyle\sum_{u=1}^{N_{\text{var}}} \left[  \frac{1 }{ N_{grid}} \displaystyle\sum_{grid=1}^{N_{\text{grid}}}\frac{(U_u^{n+1} - U^n_u)^2} {(U^n_u)^2}\right]}
\end{equation}

where \( N_{\text{var}} \) is the number of variables, \( N_{\text{grid}} \) is the number of grid cells, and \( U \) is the \( u^{\text{th}} \) variable vector at time \( n \).  Examples of the converged atmospheric volume mixing ratios and thermal structure are shown in Figures \ref{fig:VMR} and \ref{fig:TP}, respectively.

\begin{figure}[ht!]
    \centering
    \includegraphics[scale=0.38, trim=2cm 0cm 0 0cm, clip]{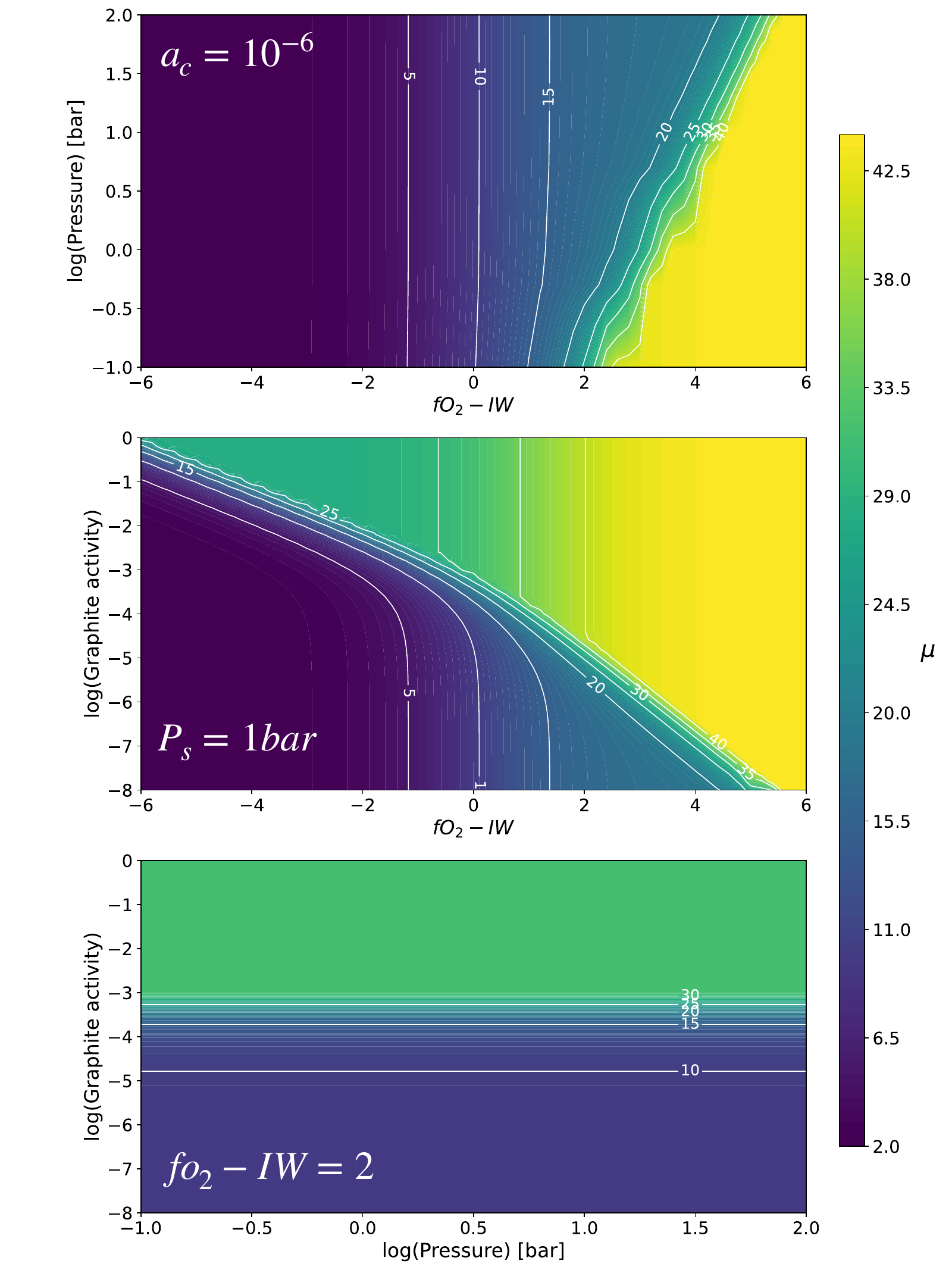}
    \caption{ Mean molecular weight (\( \mu \)) of the outgassed atmosphere (in amu) as a function of varying surface pressure, oxygen fugacity (\( f_{\mathrm{O}_2} \)), and graphite activity (\( a_{\mathrm{C}} \)). Each panel isolates the effect of one parameter by holding it fixed while allowing the other two to vary, as indicated in each panel. The value of \( \mu \) shown corresponds to mean molecular weight of the volatile partition at a given degassing rate. This figure illustrates how redox state, pressure, and carbon content influence the dominant molecular species and therefore control the overall atmospheric composition.} 
    \label{fig:graphite}
\end{figure}

In Table \ref{table:mean_molecular_weight_amu}, the simulation grid is presented in the oxygen fugacity and surface pressure space for \( a_c = 10^{-6} \). We choose \( fO_2 - IW \) values ranging from \(-6\) to \(5.0\), where IW represents the pressure- and temperature-dependent iron-wüstite buffer for oxygen fugacity \cite{Tian2024}, and \( \log f_{\rm O_2} - \mbox{IW} \) is the offset from this reference buffer in \( \log_{10} \) units. We choose surface pressures ranging from \( 0.1 \) bar to \( 100 \) bar. In Table \ref{table:mean_molecular_weight_amu}, we present the mean molecular weight (in amu) of the total atmospheric mass for each steady-state solution, where  
\[
\bar{\mu} = \sum_{i} x_i \mu_i
\]  
accounts for all atmospheric layers. We note a smooth logarithmic dependence of the mean molecular weight,  $\bar{\mu} \sim \text{log}(fO_2)$, transitioning from values near those of nebular compositions in highly reduced atmospheres to values resembling those typical of rocky solar system planets. In Figure \ref{fig:graphite}, the mean molecular weight is presented for a combination of parameters. Oxygen fugacity versus surface pressure and graphite activity \( a_c \) . Additionally, the graphite activity as a fucntion of surface pressure. We visualize those parameters to allow for intercomparison of their effect on the outgassed  mean molecular weight.  Surface pressure affects the outgassed mean molecular weight only in oxidized interiors, where \( [fO_2 - IW] \gtrsim 1 \), which generally decreases with pressure. In contrast, changes in graphite activity affect the mean molecular weight across all redox conditions except for highly reduced atmospheres with graphite activity \( \lesssim 10^{-5} \). In the following section, we highlight the importance of mean molecular weight in determining atmospheric stability. All examined atmospheric scenarios demonstrate that, given sufficient interior volatile inventories, the planet can generate a wide range of secondary atmospheres, in contrast to the solar system's dichotomy of high-mean-molecular-weight atmospheres for rocky planets and low-mean-molecular-weight atmospheres for gas giants. Once we obtain an atmosphere in steady-state conditions, we generate synthetic spectra for emission viewing geometries. Section \ref{sec:MOCK} provides details on this process. Finally, we convert our synthetic spectra into observational predictions for remote sensing observations with JWST to evaluate the feasibility of future detections under various atmospheric scenarios and to identify which spectral features could theoretically be detected using emission spectroscopy. For this transformation, we utilize the PANDEXO\footnote{\url{https://exoctk.stsci.edu/pandexo/}} version software, developed to optimize JWST and Hubble Space Telescope (HST) observations, as described in \cite{Batalha2017}.

\begin{table}[h!]
\centering
\caption{Mean molecular weight (amu) of the total atmospheric mass in the simulation grid.}
\begin{tabular}{c|cccccc}
\hline
\hline
 & \multicolumn{6}{c}{Surface Pressure [Bar]} \\  
$fO_2$-IW & 0.1  & 1.0  & 5.0  & 10.0  & 50.0  & 100.0 \\ 
\hline
-6.0 & 2.02 & 2.02 & 2.02 & 2.02 & 2.02 & 2.02 \\
-5.0 & 2.02 & 2.02 & 2.02 & 2.02 & 2.02 & 2.02 \\
-4.0 & 2.03 & 2.03 & 2.03 & 2.03 & 2.03 & 2.03 \\
-3.5 & --- & --- & --- & --- & 2.04 & --- \\
-3.0 & 2.05 & 2.05 & 2.07 & 2.05 & 2.06 & 2.06 \\
-2.5 & --- & --- & --- & 2.10 & --- & 2.10 \\
-2.0 & 2.12 & 2.12 & 2.17 & 2.17 & 2.17 & 2.17 \\
-1.0 & 2.45 & 2.46 & 2.49 & 2.46 & 2.46 & 2.49 \\
0.0 & 3.33 & 3.41 & 3.42 & 3.43 & 3.44 & 3.45 \\
0.5 & --- & 4.33 & --- & --- & --- & --- \\
1.0 & 5.81 & 15.48 & 5.76 & 5.83 & 5.86 & 5.80 \\
2.0 & 7.55 & 8.23 & 10.38 & 10.45 & 10.25 & 10.35 \\
2.5 & --- & --- & 10.33 & --- & --- & --- \\
3.0 & 12.19 & 5.79 & 14.68 & 12.18 & 12.24 & 14.06 \\
3.5 & 14.83 & --- & 14.49 & 13.82 & 15.79 & 13.81 \\
4.0 & 18.23 & 19.24 & 19.62 & 16.32 & 17.20 & --- \\
4.5 & --- & --- & --- & 33.82 & 15.89 & 18.22 \\
5.0 & --- & --- & --- & --- & 16.55 & --- \\
\hline
\end{tabular}
\label{table:mean_molecular_weight_amu}
\end{table}

\subsection{Surface albedo}\label{sec:albedo}

\begin{figure}[ht!]
    \centering
    \includegraphics[scale=0.5, trim=3.7cm 0cm 5cm 0cm, clip]{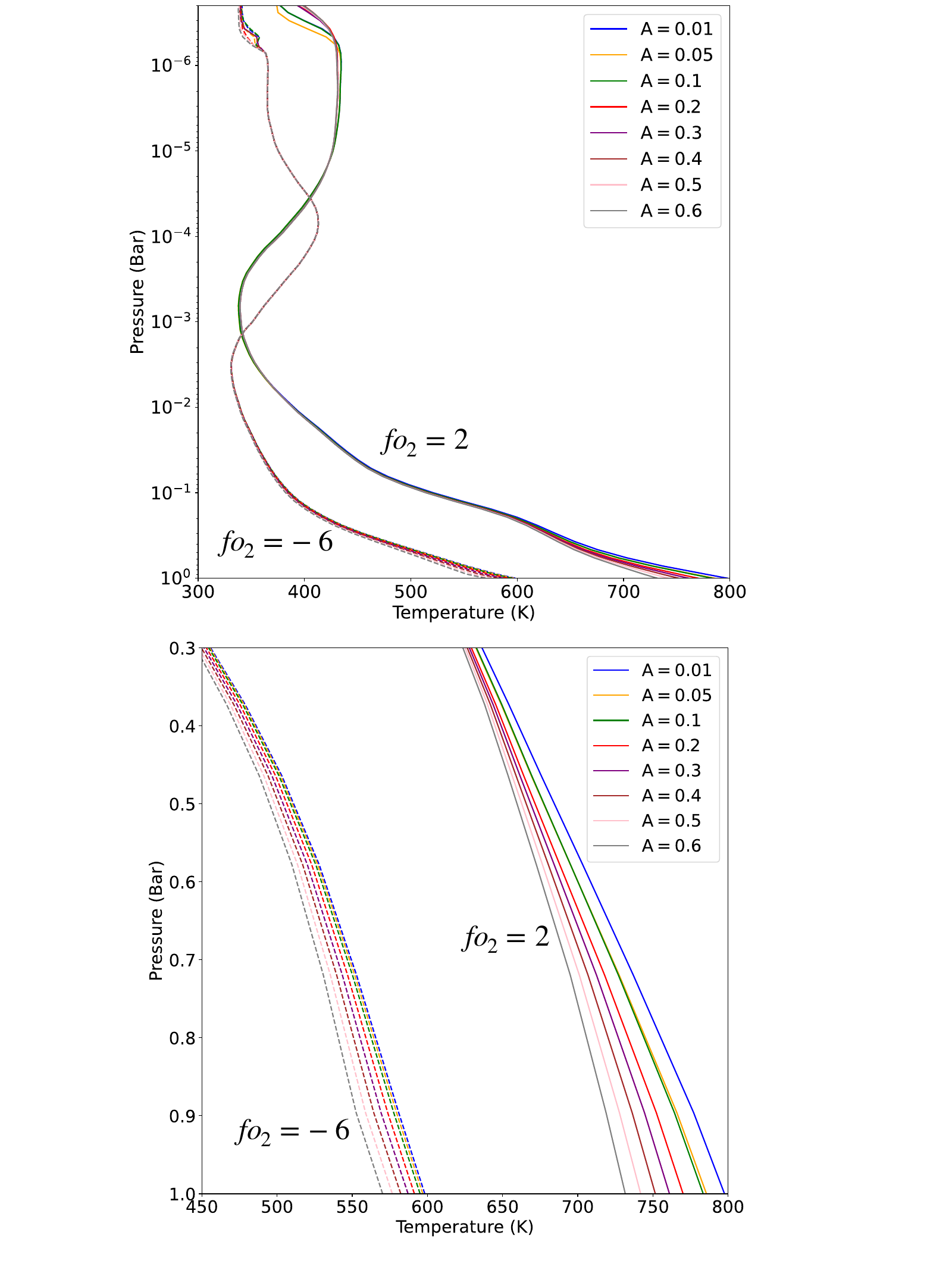}
    \caption{Top:Vertical temperature profile for a reduced and oxidized atmosphere at surface pressure of 1 bar and $a_c=10^{-6}$. We show simulations for surface albedo ranging from 0.01 to 0.6. Bottom: Detail from the top plot, focusing on the layers close to the surface.
 }
    \label{fig:albedo}
\end{figure}

For our baseline simulations, we adopted a surface albedo of 0.1, consistent with volcanic material expected for LP~791-18~d, given its anticipated high volcanic activity. To assess the sensitivity of our results to this assumption, we performed a series of simulations with albedo values ranging from 0.01 to 0.6 for both highly reduced and highly oxidized atmospheres. We explored how variations in surface albedo affect the atmospheric thermal structure, considering this broad range of plausible values representative of rocky exoplanet surfaces. Figure~\ref{fig:albedo} shows the resulting temperature–pressure profiles for both redox scenarios, assuming a surface pressure of 1~bar and a fixed graphite activity of $a_c = 10^{-6}$. The greatest sensitivity to albedo occurs in the atmospheric layers closest to the surface. In particular, surface-near temperatures vary from approximately 700 to 800~K for the oxidized case and from 550 to 600~K for the reduced one as albedo increases. However, at higher altitudes, the temperature profiles converge regardless of surface reflectivity.

While changes in Bond albedo directly affect the net stellar energy absorbed by the planet, the resulting impact on the atmospheric temperature profile is modulated by the presence of greenhouse gases. In particular, absorption and re-radiation throughout the atmosphere buffer the thermal response to increased stellar flux. In our simulations, although lowering the albedo from 0.6 to 0.01 results in a $\sim$25\% increase in equilibrium temperature (as expected from $T_{\mathrm{eq}} \propto (1 - A_b)^{0.25}$), the temperature difference is more pronounced at deeper atmospheric layers. The upper atmosphere, where radiative cooling and shortwave absorption dominate, remains comparatively less sensitive to surface albedo. This is because the greenhouse effect redistributes the incoming energy vertically, and the deposition of stellar radiation does not shift substantially upward with decreasing albedo. The upper atmospheric structure appears relatively stable across the albedo range explored here.

As a result, the impact on the thermal emission spectrum is found to be negligible—at least within the spectral bands considered in this study. This is because the photospheric pressure of those bands,  $P_{\rm photosphere} \sim g / \kappa$, lies at relatively low pressures ($\lesssim$0.1~bar),  above the region where albedo makes a significant influence.

\section{Atmospheric stability} \label{sec:stability}

All planetary atmospheres undergo escape, losing species to space, with mass loss rates varying according to planetary parameters such as gravity, atmospheric composition, and temperature structure, as well as the nature of the interaction with the stellar environment. Generally, we can distinguish two main escape regimes. The first is the planetary wind regime, which can sculpt the atmosphere and define its structure, or even its very existence. This regime involves a hydrodynamic flow where gas escapes in bulk to space \citep{Watson1981,Tian2005,Owen2019,Gronoff2020}. The second regime occurs through Jeans escape and non-thermal processes, where species escape due to their ballistic velocities exceeding the escape velocity above the exobase (the non-collisional upper layer of the atmosphere) in a direction opposite to the planet \citep[see, e.g.,][]{Shizgal1996}. In the latter case, the thermospheric part of the atmosphere, the region where the energetic portion of the stellar spectrum is deposited, can be considered hydrostatic. Here, the incoming energy is balanced by thermal conduction to the lower layers of the atmosphere, where molecular emissions radiate the energy back to space. In the non-hydrostatic scenario, however, the atmosphere balances excess incoming energy mainly through \( PdV \) work, leading to a planetary wind regime similar to Parker's wind mechanism for stellar winds \citep{Parker1958}. In our simulations, we examine the stability of these escape scenarios across the parameter space grid by applying the following methodology. We first determine the steady-state composition and thermal structure of the background atmosphere down to an atmospheric pressure of \( 10^{-8} \) bar. We then numerically define the \( R_{\text{xuv}} [\mathrm{ m}] \) radius, which corresponds to the atmospheric layer where the optical depth of hard \( XUV \) radiation reaches unity, \( \tau_{\text{xuv}} \approx 1 \), and assume that this is the layer where \( XUV \) energy is deposited. Next, we assume that the system is in an isothermal steady-state wind, a valid assumption if radiative processes occur on timescales shorter than dynamical processes, allowing the gas to regulate at a constant temperature. The spherically symmetric flow is described by \citet{Lamers1999} and follows the equation:

\begin{equation}
\dot{M} = 4 \pi r^2 \rho(r) v(r)
\label{eq:continuity}
\end{equation}
Is the continuity equation of the flow, and the momentum equation:

\begin{equation}
v \frac{dv}{dr} + \frac{1}{\rho} \frac{dp}{dr} + \frac{GM}{r^2} = 0
\end{equation}

The energy equation is to be $T(r) = T_0$ for an isothermal flow, which gives us the following solutions for the so-called critical point where the sonic velocity of the flow is achieved. 
\begin{equation}
v_s = \sqrt{\frac{kT_0}{\mu m_H}}
\end{equation}

$\mu m_H$ is the mean molecular weight in amu. The sonic radius is found by:

\begin{equation}
r_s = \frac{GM_\text{p}}{2v_s^2}
\end{equation}

The velocity profile of the wind is then given by:

\begin{equation}
\frac{v(r)}{v_s} \exp\left[-\frac{v^2(r)}{2v_s^2}\right] = \left(\frac{r_s}{r}\right)^2 \exp\left(-\frac{2r_s}{r} + \frac{3}{2}\right)
\label{eq:velocity}
\end{equation}

By combining \ref{eq:continuity} and \ref{eq:velocity} we find the density profile:
\begin{equation}
\frac{n(r)}{n_s} =  \exp\left[\frac{2 r_s}{r} -\frac{3}{2}- \frac{v^2(r)}{2v^2_s}  \right]
\end{equation}

To achieve closure for the previous set of equations and solve for the temperature \( T_0 \), we assume an energy-limited constant flow following (e.g., \citet{Sanz-Forcada2011}):

\begin{equation}
\dot{M} = \frac{\eta \pi F_{\text{XUV}} R_{xuv}^3}{G M_p}
\label{eq:mass_loss}
\end{equation}
where \( \dot{M} \) is the atmospheric escape rate [kg/s], \( \eta \) is the efficiency of the escape process, \( F_{\text{XUV}} \) is the XUV flux incident on the planet, \( R_p \) is the planetary radius, \( M_p \) is the planetary mass, and \( G \) is the gravitational constant.  The efficiency of the escape process, which can range from 0.01 to 0.6 depending on atmospheric composition and irradiation conditions \citep[e.g.,][]{Tian2005, Shematovich2014, Owen2019, Watson1981, Yelle2004, Salz2016}, is assumed to be $\eta = 0.1$. This conservative value is appropriate for secondary atmospheres, typically enriched in heavier molecules and atoms, where efficient radiative cooling processes limit the energy available for escape ( \citet{Johnstone2018eta}, \citet{Tian2005} ). 

We estimate the \( \text{XUV} \) flux reaching LP 791-18\,d based on the spectral type and the age of the star as $F_{\text{XUV}} \approx 0.1 \text{ [W/m}^2 \text{/s]}$ following:

\begin{equation}
\frac{L_{XUV}} {L_{\star}} = 
\begin{cases} 
10^{-3.6}, & t < t_{\text{sat}} \\
10^{-3.6} \left( \frac{t}{t_{\text{sat}}} \right)^{-\alpha}, & t \geq t_{\text{sat}}
\end{cases}
\end{equation}

\begin{figure}[ht!]
    \centering
    \includegraphics[scale=0.57, trim=0cm 0cm 0cm 0cm, clip]{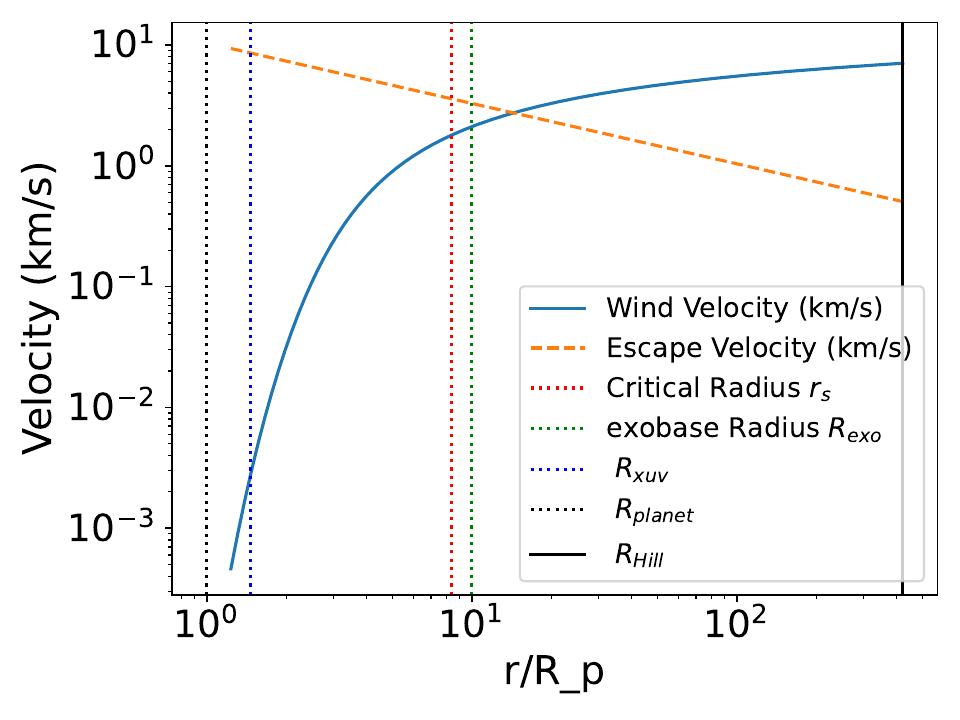}
    \caption{Wind velocity profile as a function of radius for a surface pressure of \( 1 \) bar and \( f O_2 -IW = 0 \). The blue solid line represents the wind velocity profile as a function of radius, the green dashed line indicates the exobase estimation, and the orange dashed line shows the escape velocity, which decreases as \( \sim r^{-2} \). The red dashed line marks the critical radius where the sonic velocity of the flow is reached, while the black solid line represents the Hill radius, beyond which material is no longer gravitationally bound to the planet. In this example, a planetary wind regime is established since \( r_s \leq R_{\text{exo}} \).}
    \label{fig:stability}
\end{figure}

\begin{figure}[ht!]
    \centering
    \includegraphics[scale=0.37, trim=1.cm 0cm 1cm 0cm, clip]{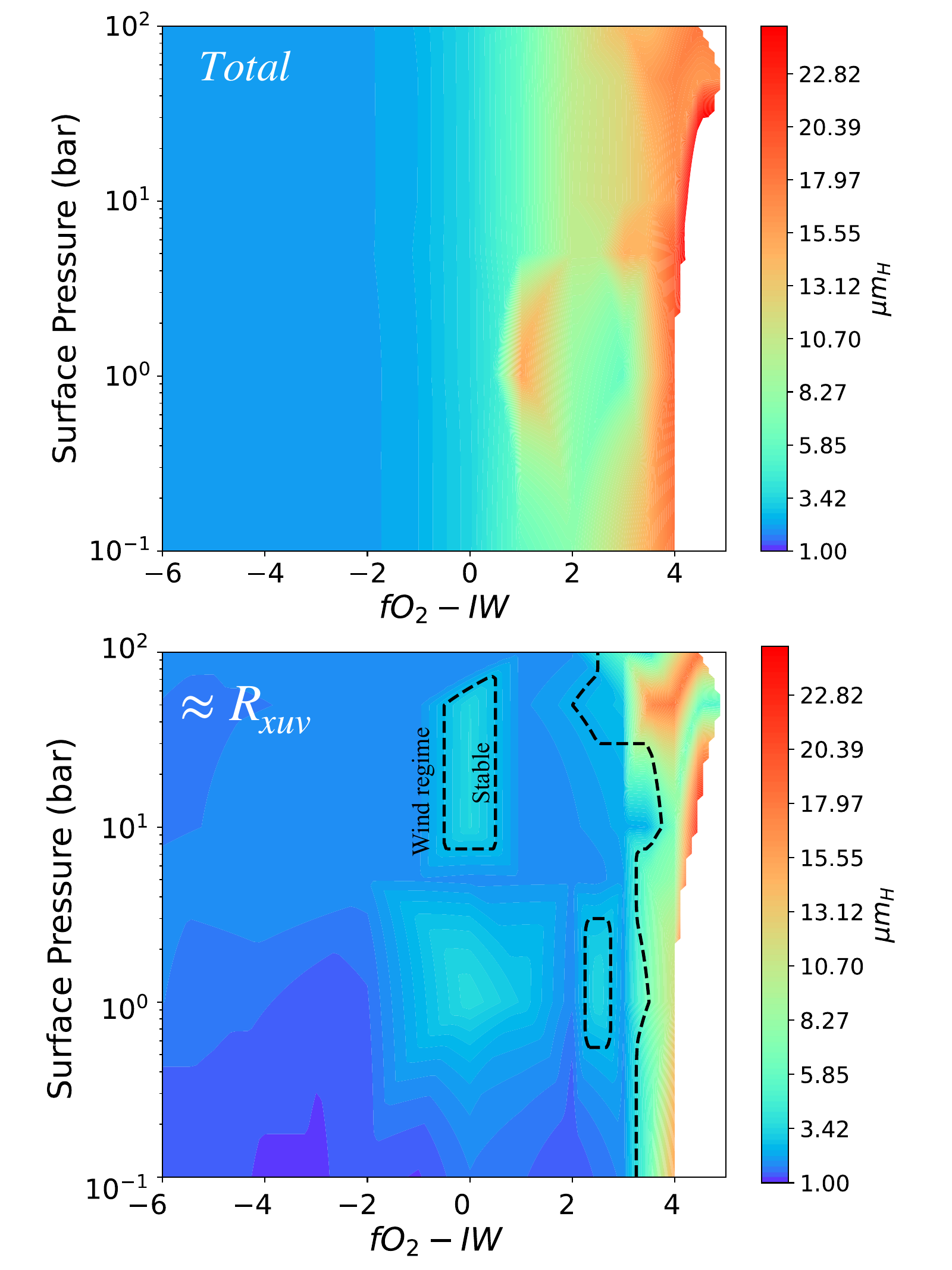}
    \caption{Mean molecular weight across the simulation parameter grid space. Top: Total atmospheric mass mean molecular weight (in amu), calculated once the simulations reach steady state. Bottom: Mean molecular weight at \( R_{\text{xuv}} \), the region where the energetic portion of the stellar spectrum is deposited, forming the base of the wind. The thick dashed black line separates the parameter space between the wind regime and stable thermospheres.  
}
    \label{fig:mean_weight}
\end{figure}

For \( L_{\star} \approx L_{\odot} \times \left(M_{\star}/M_{\odot} \right)^{4} \), \( t_{\text{sat}} = 100 \) My, the saturation interval of high-energy flux (\citet{Vilhu1987}, \citet{Wright2011}), \( t \approx 500 \) Myr, the estimated age of LP 791-18 (\citet{Reiners2009}, \citet{Newton2018}), and \( \alpha = 0.86 \) for the full XUV luminosity (see \citet{King2021}, their Figure 1), these estimations align with previous studies. We compare our results with \citet{Ramirez2014}, \citet{Johnstone2015}, \citet{Johnstone2021}, and \citet{Richey-Yowell2023} to extract \( F_{\text{XUV}} \) based on our target's characteristics. The exobase, \( R_{\text{exo}} \), is approximated by setting \( l/H(r) \approx 1 \), where  $l = \frac{1}{n \sigma_{\text{col}}}$ is the mean free path, with \( \sigma_{\text{col}} = 10^{-15} \) cm\(^2\) as the collision cross-section (\citet{Chubb2024}) and \( H(r) \) as the local scale height. The stability criterion is then defined at the supersonic point of the flow. In scenarios where the flow becomes supersonic below the exobase, we consider a planetary wind regime with mass loss rates given by \ref{eq:mass_loss}. For cases where the critical point is reached in the collisionless environment, the wind cannot be formed, and we assume a stable thermosphere. An example of this analysis is shown in Figure \ref{fig:stability} for a scenario with a surface pressure of \( 1 \) bar and oxygen fugacity of \( 0 \). We plot the velocity profile of an isothermal wind, where the temperature is solved from the system of equations presented in this section. 
The implications of these calculations for the long-term evolution of the planet's atmosphere are discussed in Section \ref{sec:discussion}. In scenarios classified as stable (i.e., not experiencing hydrodynamic escape), atmospheric loss occurs primarily through Jeans escape and non-thermal processes. We estimate these to contribute at most $\sim 1\text{--}3 \mathrm{kg,s^{-1}}$. At such low escape rates, and assuming a typical volatile inventory for a terrestrial planet, the atmosphere could persist for timescales exceeding $\sim$1Gyr. The upper panel of Figure \ref{fig:mean_weight} shows the steady-state mean molecular weight of the total atmospheric mass across the explored parameter space. The lower panel displays the mean molecular weight near the $R_{xuv}$ layer, along with the region of parameter space where atmospheric stability is achieved.

\section{Thin atmosphere to airless case} \label{subsec:bare_rock}

Following the atmospheric stability analysis in the previous section, we must consider the possibility of volatile saturation in LP 791-18\,d due to continuous hydrodynamic escape throughout the planet's lifespan. This requires accounting for the scenario of an airless rocky planet or one with a very thin atmosphere (e.g., \( <0.01 \) bar), where there is no significant heat redistribution. In such cases, the observational manifestation of a so-called bare-rock planet depends on the emissivity and reflectance of the surface material. At high surface temperatures ($\approx$1600 K), the planetary surface becomes molten and the near-infrared spectra are largely dominated by blackbody radiation. This behavior is supported from lava flow Earth observations, including laboratory experiments and in situ measurements in Hawaii (e.g., \citet{Flynn1992}, \citet{Lombardo2020}).  However, at lower temperatures, the crust interacts with electromagnetic radiation through a combination of reflection, absorption, scattering, and emission, which depends on the geometry of the interaction, the physical characteristics of the surface, the radiation wavelength, and micro-crystal formation of the material (\citet{Hapke1993,Shkuratov2001,Nelson2000}, ). 

To simulate this interaction and generate synthetic spectra, the optical properties of the surface (e.g., single-scattering albedo) are required. Even for Solar System objects, determining unique photometric parameters from disk-integrated observations alone is challenging, even when wide-phase-angle coverage is available (\citet{Domingue1991}). Thus, we must approach the problem inversely, particularly for exoplanet observations. 

Since there is no explicit guiding parameter for exploring the parameter space, unlike in atmospheric composition studies where oxygen fugacity serves as a reference, we must infer composition based on our knowledge of Earth, Solar System volcanic activity, and other analogs. Therefore, we assume a surface material and use laboratory measurements of each material to extract its optical properties. 

Reflected light depends on the geometry of the system, i.e., phase angle, latitude, and longitude of the surface area, but also on the optical properties of the surface material. Since reflected light and thermal emission may overlap, a unified treatment is required. To address this, we numerically generate a mesh-grid disk where each surface unit is assigned a set of angles, namely, the incident angle \( i \), the emergent angle \( e \), and the phase angle \( g \). The geometric transformation relations between the star-planet system and the system-observer are given by:

\begin{equation}
\mu_0 \equiv \cos i = \cos \theta \cos(\alpha - \phi)
\end{equation}
\begin{equation}
\mu \equiv \cos e = \cos \theta \cos \phi
\end{equation}
\begin{equation}
g = \alpha
\end{equation}

 The total incoming light to the observer from the whole disc can then be calculated following  \citet{Shkuratov2001} and \citet{Hu2012} as:
\begin{equation}
\frac{F_p}{F_*} = \frac{1}{F_{\text{inc}}} \int_{-\frac{\pi}{2}}^{\frac{\pi}{2}} \int_{-\frac{\pi}{2}}^{\frac{\pi}{2}} I_p(\theta, \phi) \cos^2\theta \cos\phi \, d\theta d\phi \times \left( \frac{R_p}{D_p} \right)^2
\label{eq:total}
\end{equation}

\begin{table}[h!]
\centering
\caption{Composition of assumed surface material cases.}
\begin{tabular}{c|cccccc}
\hline
\hline
\textbf{Surface} & \textbf{Si} & \textbf{O} & \textbf{Fe} & \textbf{Mg} & \textbf{Al} & \textbf{Ca/Na} \\ \hline
Metal-rich        & 0.0         & 0.1        & 0.9         & 0.0         & 0.0         & 0.0            \\
Ultramafic        & 0.19        & 0.35       & 0.12        & 0.34        & 0.0         & 0.0            \\
Feldspathic       & 0.5         & 0.3        & 0.0         & 0.0         & 0.15        & 0.05           \\
Basaltic          & 0.5         & 0.3        & 0.06        & 0.08        & 0.06        & 0.0            \\
Wustite           & 0.7         & 0.2        & 0.0         & 0.0         & 0.05        & 0.05           \\
\hline
\end{tabular}
\label{table:scenario_proportions}
\end{table}

By default, the stellar coplanar radiation originates from the direction \( (\theta = 0, \phi = \alpha) \). The incident radiation at each surface element is given by:

\begin{equation}
F_{\text{inc}} = \pi B_{\lambda}[T_*] \left( \frac{R_*}{D_p} \right)^2
\end{equation}
With \( B \) representing the wavelength-dependent stellar blackbody function, \( R_* \) the stellar radius, and \( D_p \) the star-planet distance. The total incoming radiation at each surface element is a combination of reflected, scattered, and thermally emitted radiation and is expressed in Equation \ref{eq:total} as the \( I_p \) function:
 
\begin{equation}
I_p(\theta, \phi) = I_s(\theta, \phi) + I_t(\theta, \phi)
\label{eq:sum}
\end{equation}
where:
\begin{equation}
I_s(\theta, \phi) = \frac{F_{\text{inc}} \mu_0}{\pi} r_c(\mu_0, \mu, g)
\label{eq:scattered}
\end{equation}
\begin{equation}
I_t(\theta, \phi) = \epsilon(\mu) B_{\lambda}[T(\theta, \phi)]
\label{eq:thermal}
\end{equation}

In Equations \ref{eq:scattered} and \ref{eq:thermal}, two critical parameters are introduced: \( r_c \), the bidirectional reflectance coefficient or "radiance coefficient" of solid material. This coefficient represents the brightness of a surface relative to the brightness of a Lambert surface under identical illumination (a Lambertian sphere has \( r_c = 1 \)), and it depends on the direction of both incident and scattered light. Additionally, it is influenced by the chemical composition and crystal structure of the material. Secondly, \( \epsilon \), the emissivity, quantifies the deviation of a surface material at temperature \( T_s \) from behaving as a perfect Planck emitter, given by \( B_{\lambda}[T_s] \). 

We employed  Hapke's theory of bidirectional reflectance and directional–hemispherical reflectance \citep{Hapke1981, Hapke2002} to derive the optical and photometric properties of each assumed surface composition scenario. This analytical framework accounts for single and multiple scattering in particulate surfaces and relates emissivity to reflectance through Kirchhoff’s law. We assume isotropic scatterers for multiple scattering and a simplified, cosine-based phase function for single scattering. The key optical parameter in Hapke’s model is the single-scattering albedo, which depends on the relative contributions of absorption and scattering by individual surface particles. The mean directional–hemispherical reflectance is used to compute surface emissivity for energy balance calculations in the absence of an atmosphere. For LP 791-18d, utilizing the system parameters given in \citet{Peterson2023}, the secondary eclipse lasts about $\approx 2^{\circ}$ in orbital angles.  We then have to account for the opposition surge effect—an increase in reflectance at low phase angles—through a wavelength-dependent enhancement term, $S(\lambda)$, informed by the microscopic texture and porosity of the surface material \citep{Shkuratov1999, Helfenstein1988, Gkouvelis2025}. This approach allows for a consistent treatment of both photometric and thermal properties across the spectrum. The resulting emissivity values are then used to solve for the surface temperature under radiative equilibrium, with incident flux absorbed and re-emitted according to the surface’s wavelength-dependent properties.

Space weathering effects (see, e.g., \citet{Hu2012}) will not be considered in the reflected properties of the surface material, since it has already been demonstrated by \citet{Peterson2023} and this work that the interior is highly convective, making crust replenishment likely to occur on geological timescales. 

We consider various surface material scenarios to investigate the magnitude of the bare-rock spectrum and compare it with the synthetic spectra of the atmospheric scenarios. Each surface material scenario has its own composition, where the total sum must equal one (i.e., 100\%). We simulate metal-rich, ultramafic, feldspathic, basaltic, and wüstite scenarios, which cover the spectrum from a nearly featureless thermal emission to one with strong spectral signatures—e.g., metal-rich and ultramafic, respectively. The general shape and strength of  spectral features is in agreement with the works of \citet{Hu2012}, \citet{Lyu2024} and \citet{Hammond2025} for each surface scenario. The composition of five basic scenarios is summarized in Table \ref{table:scenario_proportions}.

For each of these five scenarios, we use laboratory measurements from the United States Geological Survey Spectral Library (USGS) (\citet{Milliken2021}, \citet{Milliken2020}, \citet{Clark2007}), which are typically provided for an incidence angle of \( i = 30^\circ \), an emission angle of \( e = 0^\circ \), and a phase angle of \( g = 30^\circ \). We extract the optical parameters required for each model using the Hapke bidirectional model. To solve the inverse problem and retrieve the desired parameters, we apply the Bayesian algorithm described in Appendix B of \citet{Gkouvelis2024}, which is based on \( \chi^2 \) minimization by evaluating the Jacobians $\frac{\partial F(x)}{\partial x}$ of the forward model, which in this case is the Hapke model. The retrieved optical properties for all scenarios are the single-scattering albedo \( \omega \) , the wavelength-dependent $b$ asymmetry factor which defines the  phase function scattering parameter and  the photometric parameters relevant to the opposition effect.

\begin{figure}[ht!]
    \centering
    \includegraphics[scale=0.265, trim=1cm 0.1cm 0cm 0cm, clip]{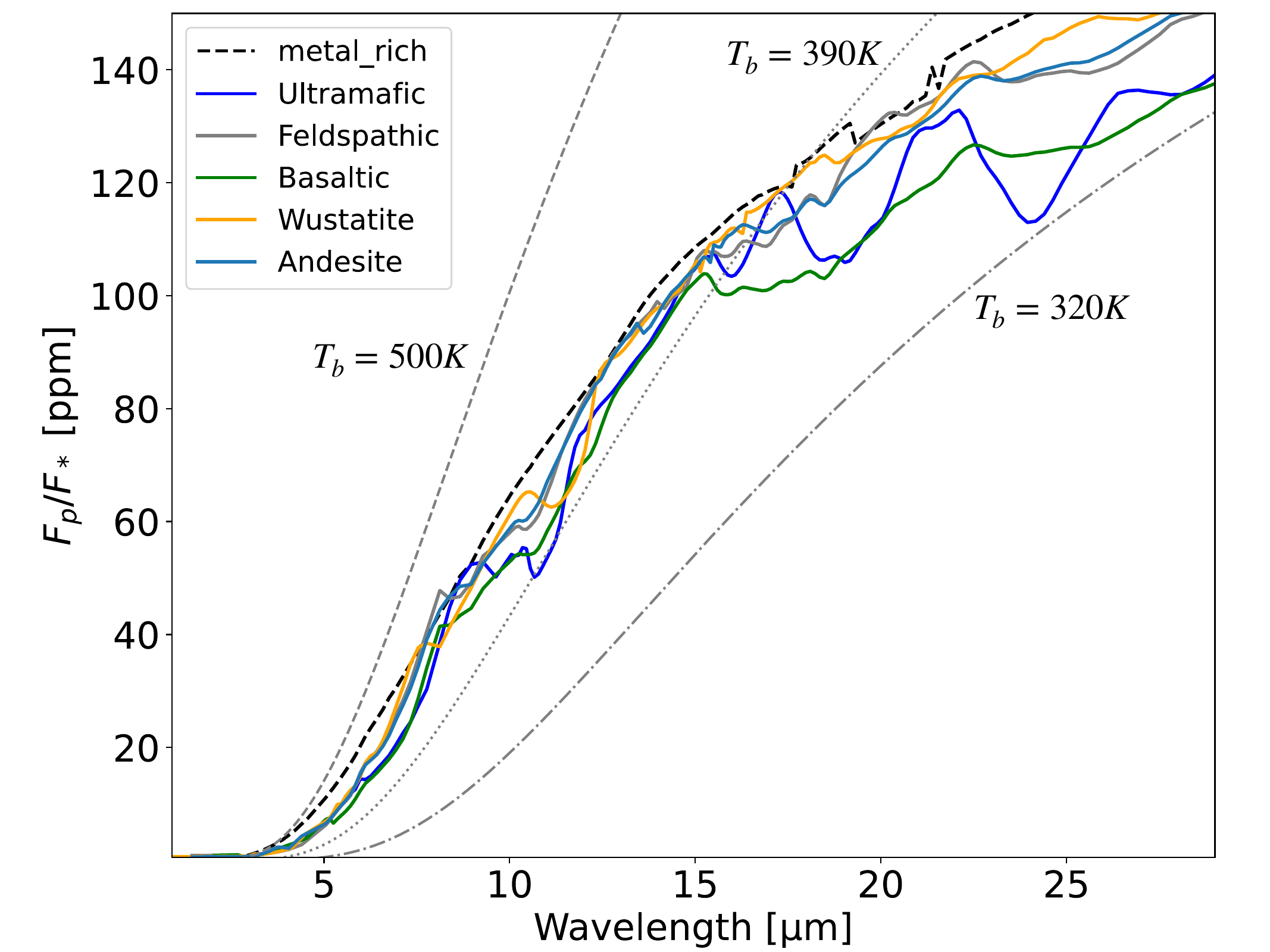}
    \caption{Airless body scenarios. Secondary eclipse depth spectra generated for various surface material scenarios without considering space weathering effects. Brightness temperature curves are overplotted as a comparative scale to highlight spectral features. }
    \label{fig:rock}
\end{figure}

The secondary eclipse depth of the system is evaluated using Equation \ref{eq:total} at a phase angle of \( g = 0^\circ \), which is standard practice in observations (e.g., when the planet approaches eclipse, see \citet{Benneke2024}) as it maximizes the contrast. The energy balance equation is evaluated numerically, with convergence solutions achieved at angular resolutions finer than \( \approx 7^\circ \) for both \( \mu \) and \( \phi \). We present the synthetic spectra of LP 791-18\,d as a bare-rock planet for the five surface scenarios, shown in Figure \ref{fig:rock}. Additionally, we overplot the isothermal brightness temperature functions \( T_b(\lambda) \) as indicators of the planet's thermal emissivity (\cite{Seager2010}). These brightness temperature functions serve as a visual guide to understanding how different wavelengths trace temperature variations for each scenario. However, it is important to note that these spectra represent disk-averaged values. In practice, this can be determined by solving for the brightness temperature \( T_b(\lambda) \) using:
\begin{equation}
 F_p=\pi B_{\lambda}(T_b)(\frac{  R_p}{D})^2 
\end{equation}
For featureless surface scenarios  the brightness temperature function \( T_b(\lambda) \) varies by only a few kelvins across the spectral range due to the smooth wavelength dependence of surface emissivity. In contrast, for surfaces with strong spectral features, \( T_b(\lambda) \) can vary by several tens of kelvins, reflecting the stronger wavelength-dependent modulation of the emergent thermal flux.

\section{Observational manifestations}\label{sec:MOCK}

In this section, we focus on observable manifestations that can be used to characterize the planet and derive physical properties, such as atmospheric chemical composition and its correlation with oxygen fugacity, which serves as an indicator of the interior state. The forthcoming JWST observations of LP 791-18\,d will target only the \( \text{CO}_2 \) \( 15\, \mathrm{ \mu m} \) strong absorption band. Therefore, we first analyzed this band based on our simulated spectra.  We demonstrate that, based on our simulations, the objectives proposed in \citet{Benneke2024} can be achieved for a limited subset of scenarios within our simulation grid space (see also the discussion section).  
We then propose alternative strategies for characterizing LP 791-18\,d, which may also serve as a general approach for studying similar Earth-size exoplanets.

\subsection{Emission} \label{emission}

Thermal emission originates directly from the planet and can be detected through its thermal contrast with the host star, expressed as:

\begin{equation}
\frac{ F_p(\lambda) }{ F_*(\lambda)  }=\frac{ f_p(\lambda) }{ f_*(\lambda)} \frac{ R_p^2 }{ R_*^2 }  \Phi(\alpha) + \left(\frac{R_p}{a}\right )^2 A_g  \Phi(\alpha)
\label{eq:emission}
\end{equation}

\begin{figure}[ht!]
    \centering
    \includegraphics[scale=0.44, trim=0.2cm 0.5cm 0cm 1cm, clip]{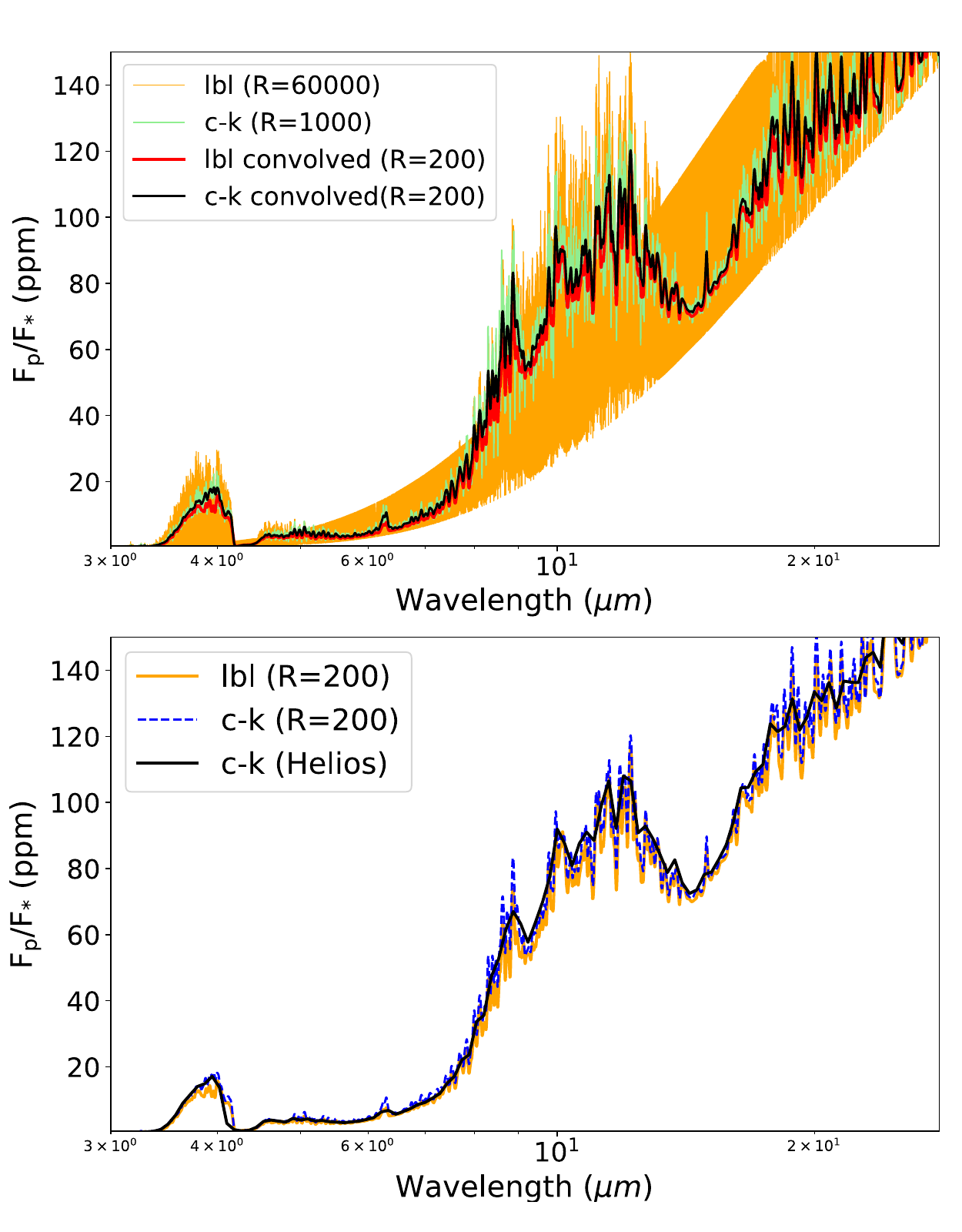}
    \caption{Thermal emission spectrum for a simulation with surface pressure of 1~bar, redox state of $\mathrm{fO}_2 - \mathrm{IW} = 2$, and $a_c = 10^{-6}$. \textit{Top:} Comparison between the line-by-line radiative transfer method at $R = 60{,}000$ and the correlated-$k$ method at $R = 1{,}000$, both convolved to $R = 200$. \textit{Bottom:} Comparison of convolved spectra from both line-by-line and correlated-$k$ methods generated using \texttt{petitRADTRANS}, along with the output spectrum from our framework. All three approaches show negligible differences when predicting photometric fluxes in the JWST MIRI broad-band filters (see text for more details).
 }
    \label{fig:high_res}
\end{figure}

Where \( \Phi(\alpha) \) is the orbital phase angle of the planet. It is defined as \( \alpha = 0 \) at primary transit, when the observer views the planet’s night side directly, and \( \alpha = 1 \) at secondary eclipse, when the planet is behind the host star. The second term represents the contribution from scattered light; more details on its derivation can be found in \citet{Seager2010}. Here, \( \alpha \) denotes the semi-major axis of the star-planet system, and \( A_g \) is the geometric albedo. As discussed in previous sections, our numerical simulations include detailed radiative transfer calculations. As a result, we obtain the outgoing radiation from the planet in 73 spectral bins, which serves as \( f_p(\lambda) \), while the incoming radiation at the top of the atmosphere, \( f_*(\lambda) \), is one of our model inputs. For the latter, we used the stellar spectrum of LP~791-18 from the MUSCLES catalog \citep{Behr2023, Wilson2025}. This allows us to directly compute the thermal emission spectrum for each of our simulations.
Our theoretical framework solves the radiative transfer using the correlated-$k$ method at relatively low spectral resolution ($R \sim 70$) for computational efficiency. To assess the impact of spectral resolution on the computed emission spectra and the resulting photometric observables, we conducted a numerical experiment to test the robustness of our approach. For a representative subset of simulations, we generated high-resolution spectra using the line-by-line mode of \texttt{petitRADTRANS}\footnote{\url{https://petitradtrans.readthedocs.io/en/latest/\#}}
\citep{Molliere2019, Molliere2020, Alei2022, Nasedkin2024} at a resolving power of $R = 60{,}000$, which were then convolved to $R = 200$. We also computed spectra using the correlated-$k$ method at $R = 1{,}000$, likewise convolved to $R = 200$. We found that the synthetic photometry derived from these higher-resolution spectra was in excellent agreement with that obtained using the 
 correlated-$k$ method as well as the original \texttt{HELIOS} output. This confirms that the spectral resolution of the \texttt{HELIOS} spectra is sufficient for producing reliable photometric predictions when integrated over the broad MIRI filter bandpasses ($R \sim 10 - 20 $). A comparison of the results from \texttt{HELIOS} and \texttt{petitRADTRANS} is shown in Figure \ref{fig:high_res}. 
 
\begin{figure}[ht!]
    \centering
    \includegraphics[scale=0.48, trim=0.4cm 0.45cm 0cm 0cm, clip]{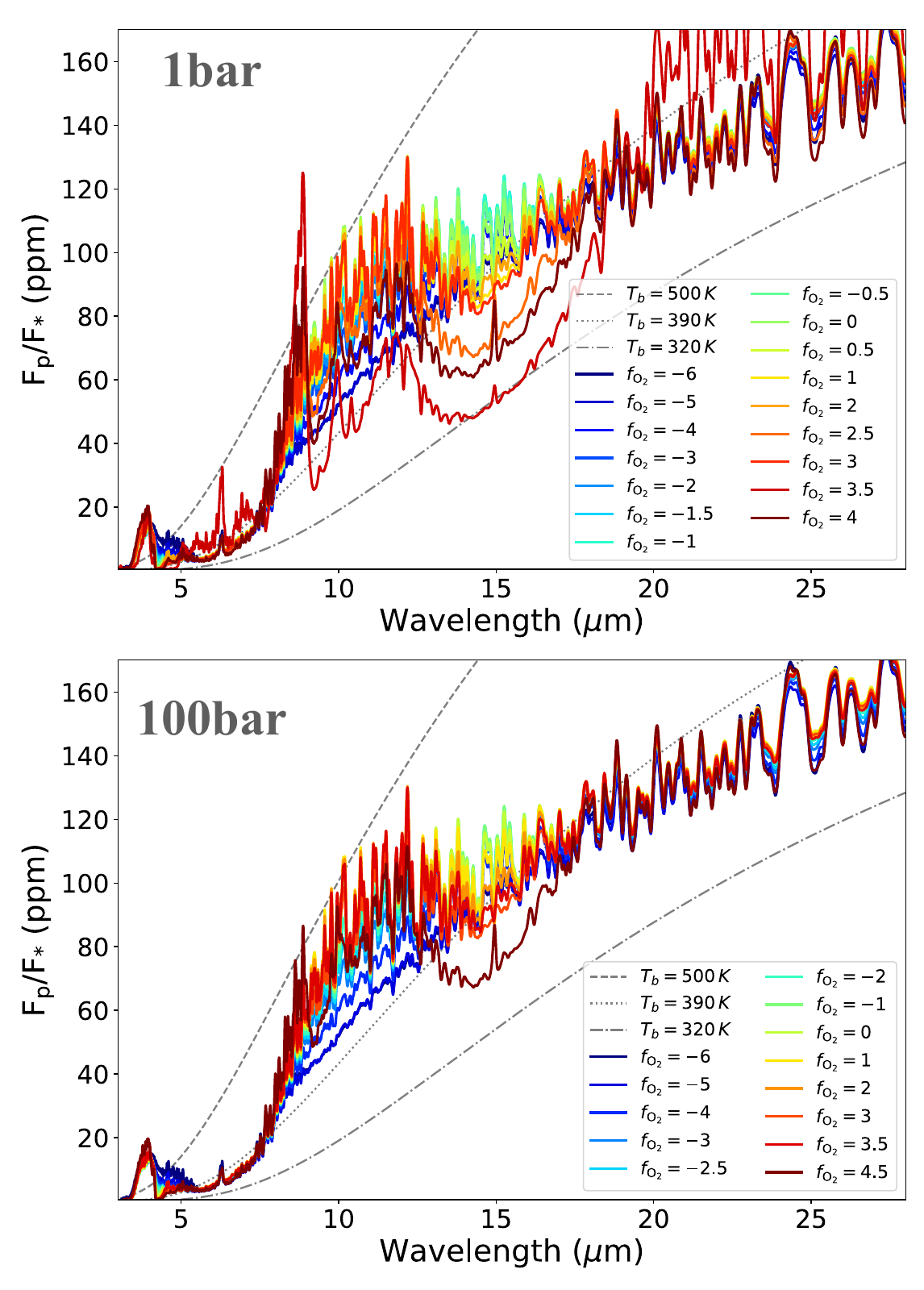}
    \caption{Thermal emission spectra of $1\text{bar}$  and  $100\text{bar}$ surface pressure simulations across various oxygen fugacities at $a_c=10^{-6}$.  The CO$_2$ $15\, \mathrm{\mu m}$ absorption band becomes prominent for oxidized atmospheres but it is subdued for thick atmospheres.  }
    \label{fig:emission}
\end{figure}

 The photometric results presented in this work are based on synthetic spectra generated with \texttt{petitRADTRANS} using the correlated-$k$ method at intermediate resolution ($R = 1{,}000$), subsequently convolved to $R = 200$. The HELIOS code was used exclusively to compute the radiative–convective equilibrium structure of the atmosphere and achieve steady-state temperature profiles. These profiles were then passed to \texttt{petitRADTRANS} to calculate self-consistent emission spectra. While \texttt{petitRADTRANS} was used here for its flexibility and higher spectral resolution capabilities, either tool could be used to produce synthetic spectra, and the choice does not impact the scientific conclusions drawn in this work.
 An example is presented in figure \ref{fig:emission} for different simulated scenarios. We also overplot the isothermal brightness temperature, \( T_b(\lambda) \), as an indicator of the temperature traced at various atmospheric layers at each wavelength (\cite{Seager2010}). This follows the relation $P_{\text{photosphere}} \approx \frac{g}{\kappa}$. In the plotted example, the atmospheric layers are enclosed within a temperature range of approximately 320 to 500 K. 

Due to variations in atmospheric composition, we observe differences in the thermal emission spectra across the oxygen fugacity parameter space.  The primary spectral features of interest include:  
1) A strong absorption band around \( 15\, \mathrm{\mu m} \), which is caused by CO\(_2\) absorption and is therefore more prominent in oxidized atmospheres.  2)An enhanced continuum feature appears around \( 9\,\mu\mathrm{m} \), arising from the absence of CH\(_4\) and H\(_2\)O opacity. Under more oxidizing conditions, the depletion of these hydrogen-bearing species opens a spectral window, allowing radiation at this wavelength to escape from deeper and hotter atmospheric layers. The strength of the emission feature increases rapidly with oxygen fugacity, owing to the logarithmic nature of the \( f_{\mathrm{O}_2} \) scale. As \( f_{\mathrm{O}_2} \) crosses a critical threshold (e.g., IW\,+\,3), the atmospheric composition undergoes a sharp transition, with the near-complete loss of CH\(_4\) and H\(_2\)O. This transition enhances thermal emission at 9\,\(\mu\)m, making the feature especially prominent in that regime.

\begin{figure*}[ht!]
    \centering
    \includegraphics[scale=0.62, trim=0cm 0cm 0cm 0cm, clip]{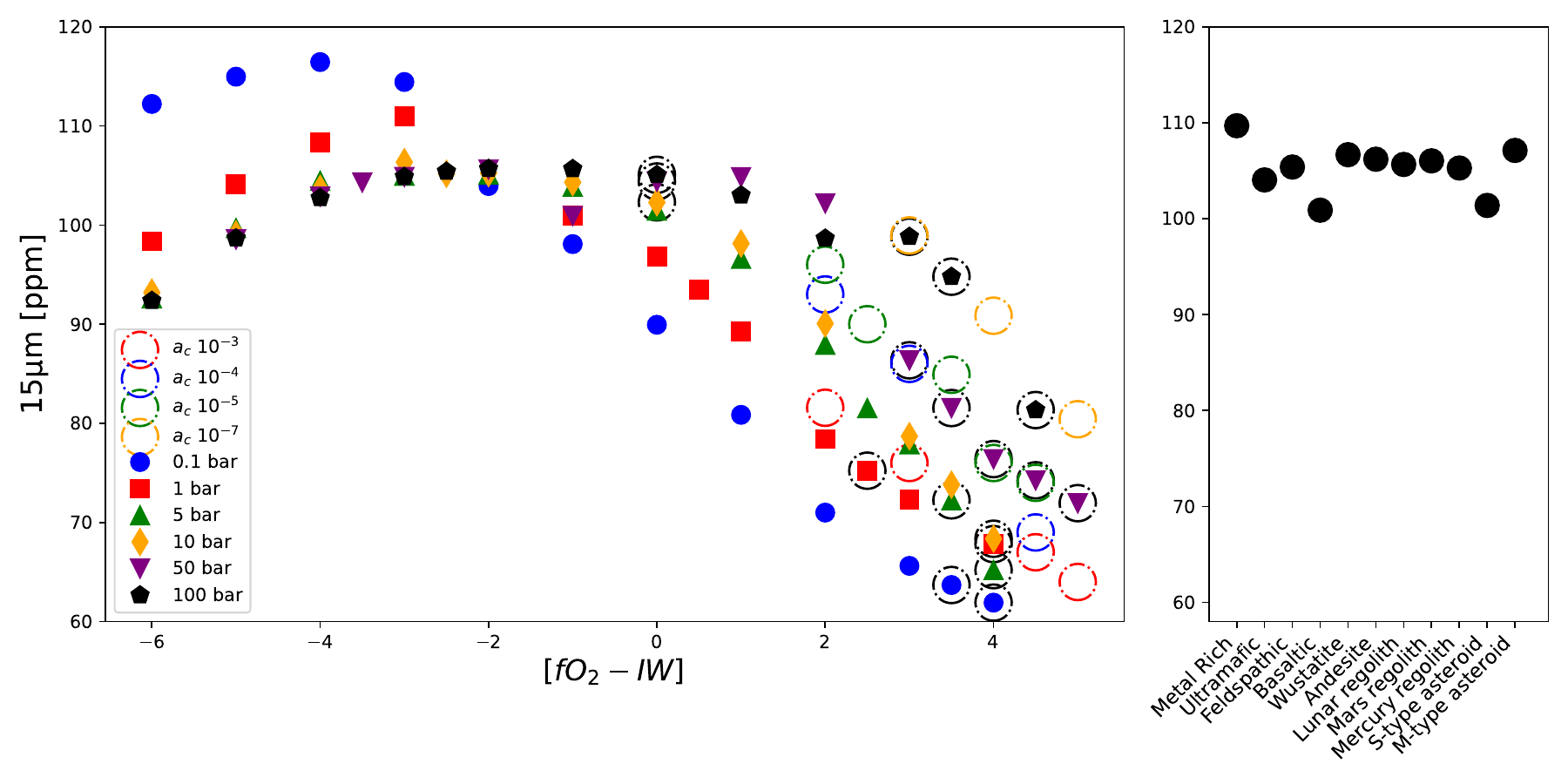}
    \caption{\textbf{Left:} $15\, \mathrm{\mu m}$ band secondary eclipse depth as a function of $fO_2$ across the surface pressure grid space with main graphite activity $a_c=10^{-6}$ . Dashed circles indicate stable atmospheres, as approximated using the methodology described in Section \ref{sec:stability}. Symbols in black indicate the stable atmospheres of all surface pressure scenarios for the main graphite activity value, while all other colors indicate the variability in graphite content for a 100 bar atmosphere. \textbf{Right:} \( 15\, \mathrm{\mu m} \) band for the bare-rock with various surface material scenarios, plotted on the same vertical scale as the left panel.  
 }
    \label{fig:15umVSFO2}
\end{figure*}

In Figure \ref{fig:15umVSFO2}, we show the variation of the \( 15\, \mathrm{\mu m}\) spectral band (where the MIRI F1500W filter’s effective wavelength is \( \lambda = 14.79\,  \mathrm{\mu m} \)) integrated over \( \pm1.5 \mu m \) to match the MIRI F1500W filter bandwidth (\citet{Greene2023}, \citet{Hammond2025}). We plot the \( 15\, \mathrm{\mu m}\) spectral band for each of our simulations, including the airless planet scenarios, using the same scale (right-hand section of Figure \ref{fig:15umVSFO2}). We find that bare-rock eclipse depths are comparable to those of reduced atmospheric scenarios, with mean differences of approximately \( 30-40 \) ppm relative to oxidized atmospheres. To assess the detectability of these scenarios with JWST’s F1500W filter, we applied our synthetic spectra to the PANDEXO software (\cite{Batalha2017}) to estimate uncertainties for each simulation. Using the observational characteristics described in \citet{Benneke2024}, we find error bars on the order of \( \approx \pm30 \) ppm, depending on the specific case within our simulation grid. \\  
The estimated level of uncertainty suggests that distinguishing an atmospheric detection from an airless case is possible, but with low confidence, and only in the case of a thin and highly oxidized atmosphere. We discuss this issue in more detail in the final section. In the following parts of this section, we explore alternative approaches beyond solely relying on \( 15\, \mathrm{\mu m} \) photometric observations.   

\begin{figure}[ht!]
    \centering
    \includegraphics[scale=0.46, trim=0.5cm 1cm 1.8cm 0cm, clip]{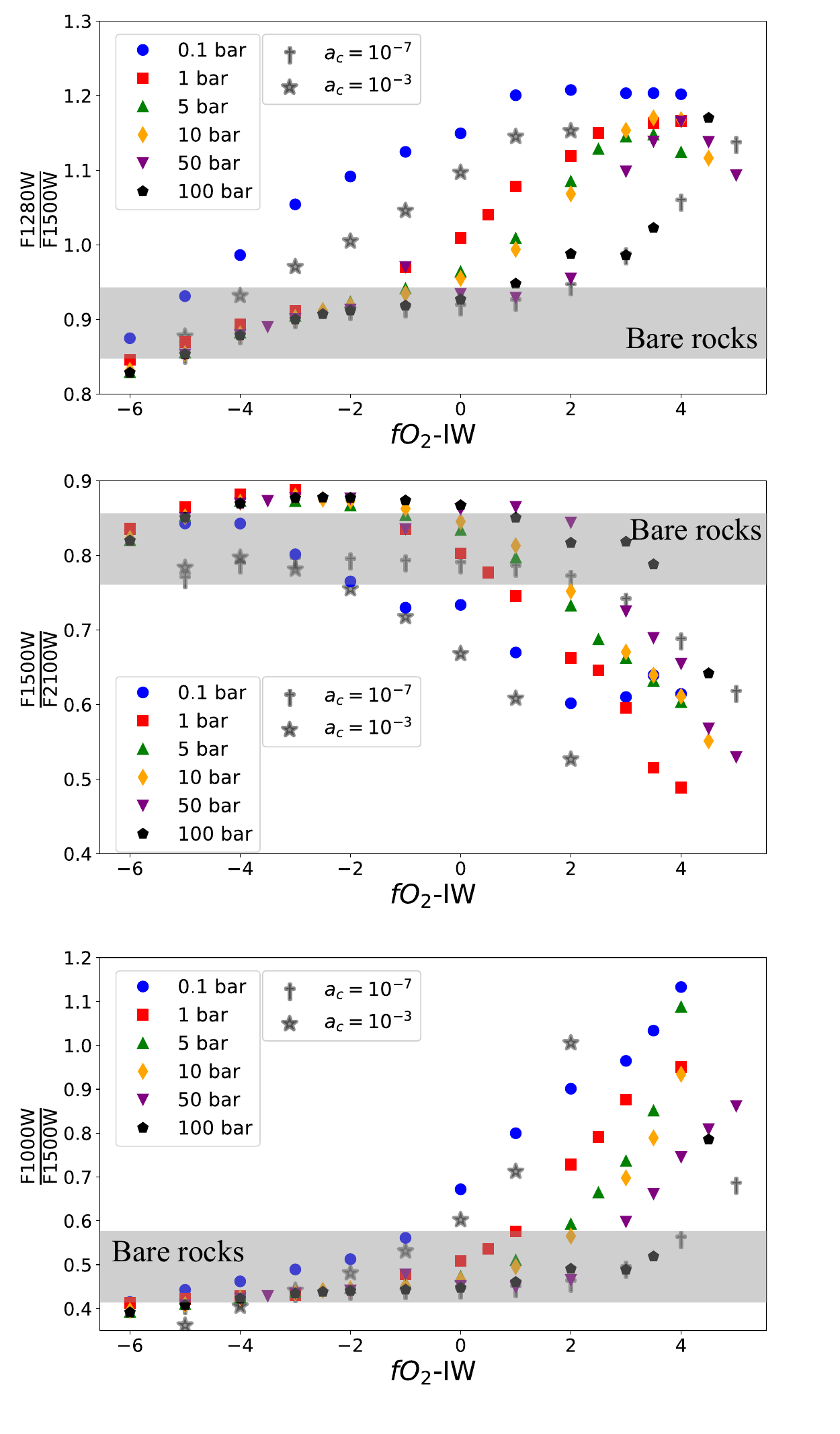}
    \caption{Photometric band ratios for various surface pressures as a function of oxygen fugacity at $a_c=10^{-6}$. We overplot the extreme values of graphite activity, $a_c$, $10^{-7}$ and $10^{-3}$ for 100$\text{bar}$ surface pressure. \textbf{Top:} Ratio of the \( 12.8\, \mathrm{\mu m} \) to \( 15\, \mathrm{\mu m} \) spectral bands. \textbf{Middle:} Ratio of the \( 15 \, \mathrm{\mu m} \) to \( 21 \, \mathrm{\mu m} \) \textbf{Bottom:} Ratio of the \( 10 \, \mathrm{\mu m}\) to \( 15 \, \mathrm{\mu m} \). Variation in surface pressure is indicated with different colored shapes.
 }
    \label{fig:indexes}
\end{figure}

Firstly, we analyzed the secondary eclipse depth for spectral bands adjacent to the \( 15\, \mathrm{\mu m} \) band. This observational approach was recently adopted by \citet{Ducrot2024}, who conducted follow-up observations of TRAPPIST-1 b in the \( 12.8\, \mathrm{\mu m} \) band using MIRI’s F1280W filter. We present the ratio of the F1280W to F1500W filters for our simulated scenarios as a function of oxygen fugacity in Figure \ref{fig:indexes} (upper plot). Furthermore, from the full list of MIRI photometric bands, we identify two bands that correlate with the \( 15\, \mathrm{\mu m} \) band: The \( 9\, \mathrm{\mu m} \) band, where CH\(_4\) exhibits strong absorption. The \( 20 \, \mathrm{\mu m} \) band, which remains largely insensitive to oxygen fugacity or surface pressure variations (at least within our numerical experiments). \\ Across all filter indices, variations of up to \( 50\% \) are observed in the oxygen fugacity space. Additionally, increasing surface pressure suppresses the proposed filter index ratios, leading to conclusions similar to those of \citet{Hammond2025}: the degeneracy between thick atmospheres and bare-rock planets remains unresolved, even when analyzing multiple photometric bands. Nevertheless, examining multiple bands provides more information than a single-band approach. A clear correlation with the redox state emerges across all photometric band indices, suggesting that, in principle, such observations could offer insights into the planet’s interior redox state.  Seeing Figure \ref{fig:indexes}, the ratio of any of those three photometric pairs could indicate a reduced or oxidized interior. For example, a  low ratio value between F1500W with F1200W or F1000W, on average, is an indicator of a reduced interior while F1500W with F2100W is the opposite. However, for LP 791-18 d, this capability remains beyond the reach of current instrumentation. 

\subsection{A JWST color-color diagram} \label{color}
\begin{figure*}[ht!]
    \centering
    \includegraphics[scale=0.45, trim=0cm 0cm 0cm 0cm, clip]{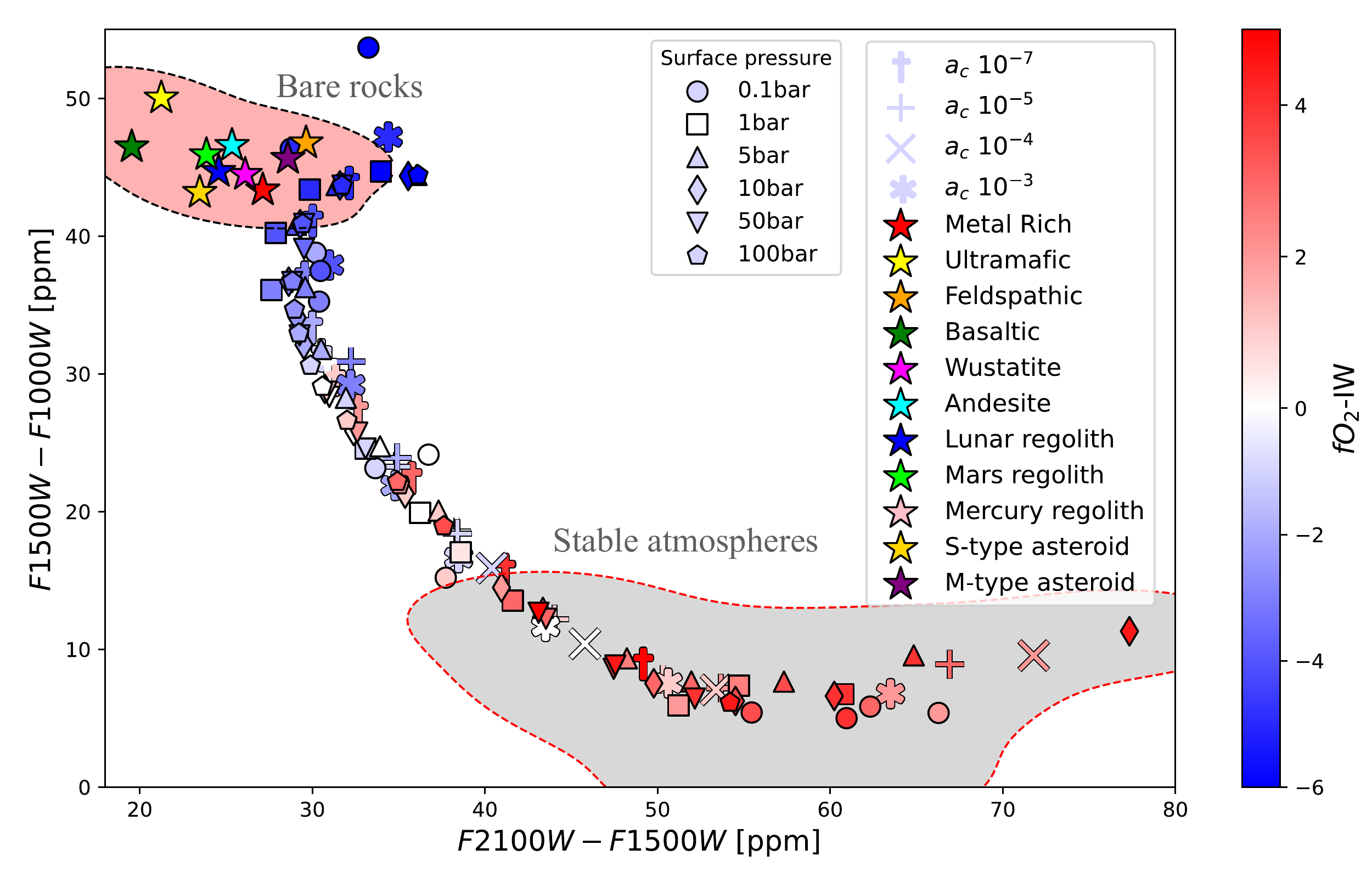}
    \caption{Proposed color-color diagram that breaks the degeneracy between bare-rock and thick oxidized stable atmospheres. The combination and names of the three employed JWST/MIRI photometric bands are shown on the axes. A set of bare-rock surface material scenarios is represented by colored stars ($\star$). The various shapes indicate values across our grid of surface pressures, with colors representing the oxygen fugacity values. Blue represents reduced conditions and red represents oxidized conditions. Finally, we overplot different graphite activity (\( a_c \)) values with different  symbols (see legend: \( \dagger \), \( + \), \( \times \), and \( * \)) at random surface pressures. The shadowed areas represent the KDE probability density of the bare-rock and stable-atmosphere clusters.  
}
    \label{fig:color}
\end{figure*}

A natural continuation of the analysis conducted so far, both in this manuscript and in the literature, is to introduce an additional photometric band and explore its potential. With three photometric bands, we can plot a color-color diagram and look for clustering of the possible scenarios. In Figure \ref{fig:color}, we extract photometric secondary eclipse depths from our simulations for the characteristics of MIRI filters F1000W, F1500W, and F2100W, which correspond to \( 10 \, \mathrm{\mu m}\), \( 15 \, \mathrm{\mu m}\), and \( 21 \, \mathrm{\mu m} \), respectively. To estimate the observational uncertainties, we used the \texttt{PANDEXO} simulator \citep{Batalha2017}, adopting the same configuration proposed for LP~791-18\,d observations by \citet{Benneke2024}, namely five visits using the F1500W filter with the SUB256 sub-array. Identical integration times were assumed for the F1000W and F2100W bands to derive corresponding error bars. We observe clustering of the bare-rock scenarios in the upper-left corner of the figure, while the more reduced scenarios nearly overlap with the bare-rock color-color space. As the interior redox state approaches more oxidized conditions, the population gradually separates. The stable atmospheres are clearly separated from the bare-rock scenarios in the color-color diagram. Both surface pressure and graphite activity effects become negligible in the diagram, meaning that, for example, a separation of stable atmospheres with \( 1 \) bar is nearly indistinguishable from \( 100 \) bar or graphite activity values ranging from \( 10^{-7} \) to \( 10^{-3} \). To visualize the density of scenarios in color–color space, we compute a two-dimensional Kernel Density Estimation (KDE) using a Gaussian kernel with bandwidth determined via Scott’s Rule. The shaded contours in Figure \ref{fig:color} correspond to the $2\sigma$ (95\%) confidence region of the KDE distribution for each population. The two point clouds are for bare rocks and the stable atmospheres for the other.

\subsection{Impact of potential photochemical hazes and clouds}\label{sec:hazes}

Although our current modeling framework does not explicitly include photochemical haze formation, we acknowledge that such aerosols may play a non-negligible role in shaping the atmospheric observables of LP~791-18\,d, particularly given the planet’s M-dwarf host star and potential for UV-driven chemistry. In analogy to Titan’s atmosphere, photochemical production of complex organic molecules could lead to high-altitude haze layers, especially in reducing or hydrocarbon-rich atmospheres. These hazes can increase the planetary albedo, reducing the dayside brightness temperature, and can also flatten or obscure infrared molecular features in emission spectra.
To investigate the impact of photochemical hazes on the emission spectra of LP~791-18\,d, we employ a  two step simplified parametric approach that captures their scattering and extinction.  First, we scale the Rayleigh scattering cross sections of gas species using the \texttt{haze\_factor}  ($H_{f}$) parameter in \texttt{petitRADTRANS} \citep{Molliere2019}. This scaling enhances the native Rayleigh opacity of the gas mixture, simulating the wavelength-dependent scattering introduced by small-particle hazes. $H_{f}$ multiplies the Rayleigh scattering opacity (in cm$^2$\,g$^{-1}$) of gas species. That is, if the baseline Rayleigh opacity is denoted by $\kappa_{\mathrm{Rayleigh}}$, the total scattering opacity becomes:
\begin{equation}
\kappa_{\mathrm{total}} = H_{f} \times  \kappa_{\mathrm{Rayleigh}} 
\end{equation}
This parameter allows a simplified treatment of enhanced atmospheric scattering due to photochemical hazes and scales with the assumed aerosol content as $\sim n_{aerosols} \times \sigma_{scat}$.  E.g. $H_{f}$ = 1: standard Rayleigh scattering from gaseous species only. $H_{f}$ $>$ 1: mimics additional scattering by sub-micron aerosol particles.  $H_{f}$ $\gg 1$  represents optically thick hazes with significant scattering at short wavelengths.  This parameterization serves as a computationally efficient proxy for the enhanced scattering from small photochemical haze particles, such as complex hydrocarbons (e.g., tholins), which can form particles with radii $\lesssim 0.5~\mu$m and introduce strong wavelength-dependent opacity in the UV to near-IR range. Additionally, we incorporate a toy model for wavelength-dependent absorption, representing Mie-like opacity  to mimic additional broadband absorption from larger aerosols, including condensate clouds,  defined as:
\begin{equation}
    \kappa_{\rm haze}(\lambda) = \kappa_0 \left( \frac{\lambda}{1~\mu\mathrm{m}} \right)^{\beta}
\end{equation}

where $\kappa_0$ is the opacity at $1~\mu$m which can vary from 0.1 to 10 $\text{cm}^2$ g$^{-1}$ depending on composition. $\beta$ is the power-law index, typically $\beta = -4$  for small haze particles  (\citet{Etangs2008}) and size $\alpha \ll \lambda $  where Rayleigh scattering dominates. For larger particles, $\alpha\sim \lambda$, for which Mie scattering dominates $\beta$   factor tends to values closer to $ -1$  .  Typical haze particles are in the range of $0.01 - 0.3 \, \mathrm{\mu m}$ \citet{Lavvas2008}, \citet{Lavvas2010} , \citet{Lavvas2019} , \citet{Gao2021}.  Nevertheless,  larger particles can be produced from coagulation and growth  (see, e.g., Figure 8 from \citet{Gao2021}). 

\begin{figure}[ht!]
    \centering
    \includegraphics[scale=0.54, trim=4cm 0cm 6cm 0cm, clip]{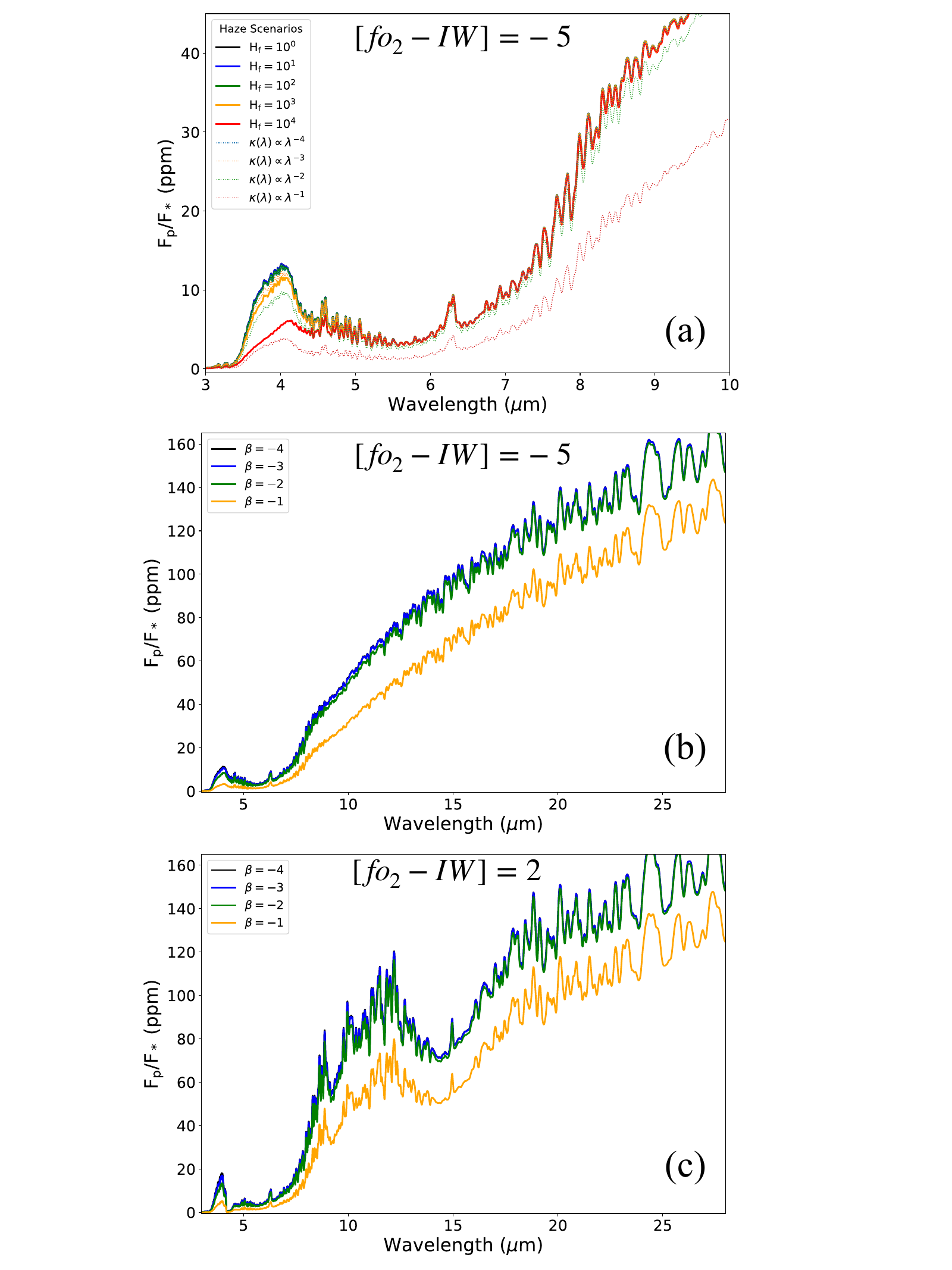}
    \caption{Effects of photochemical haze aerosols. (a) Reduced atmospheric scenario with $a_c= 10^{-6}$ and surface pressure 1bar is simulated across the haze enhancement factor $H_f$ and $\beta$. (b) Same atmosphere as (a), but the full spectral range is presented and across $\beta$ variances. (c) Oxidized atmosphere with $a_c= 10^{-6}$ and surface pressure 1bar is presented to show the effects on the $15 \, \mathrm{\mu m}$ CO$_2$ absorption band.
 }
    \label{fig:haze}
\end{figure}
This absorption-like attenuation is applied to the synthetic spectra as an exponential suppression $\exp(-\tau_{\mathrm{haze}})$ to explore the impact of Mie-type extinction.  Together, these methods allow us to approximate a broad range of possible haze and condensate cloud effects in a computationally efficient way, without explicitly modeling microphysical processes.

 In figure \ref{fig:haze} (a) we show the effect of $H_f$ across four orders of magnitude, with solid lines. 
We have performed emission spectrum calculations across a range of haze enhancement factors, from $H_{f} = 1$ (clear atmosphere) up to $10^4$ (opaque hazy atmosphere), to assess their impact on the observable planet-to-star flux contrast (see figure \ref{fig:haze}). These values encompass plausible ranges for hazy atmospheres, as observed in solar system bodies like Titan and in photochemically active exoplanets \citep{Arney2016, Gao2021}. The haze is assumed to be well-mixed and vertically distributed throughout the atmospheric pressure range.
While this approach does not model haze microphysics explicitly, it provides a computationally efficient means to explore the sensitivity of thermal emission features to scattering and extinction by photochemical aerosols. Our simulations indicate that significant haze enhancement ($H_{f} \gtrsim 10^3$) is required to appreciably mute emission features, particularly at near-infrared wavelengths. At $H_{f} \sim 10^4$  the atmosphere is saturated and the short wavelengths ($\lesssim 4 \, \mathrm{\mu m}$ ) are flattened.

\begin{figure}[ht!]
    \centering
    \includegraphics[scale=0.255, trim=0cm 0cm 0cm 2.3cm, clip]{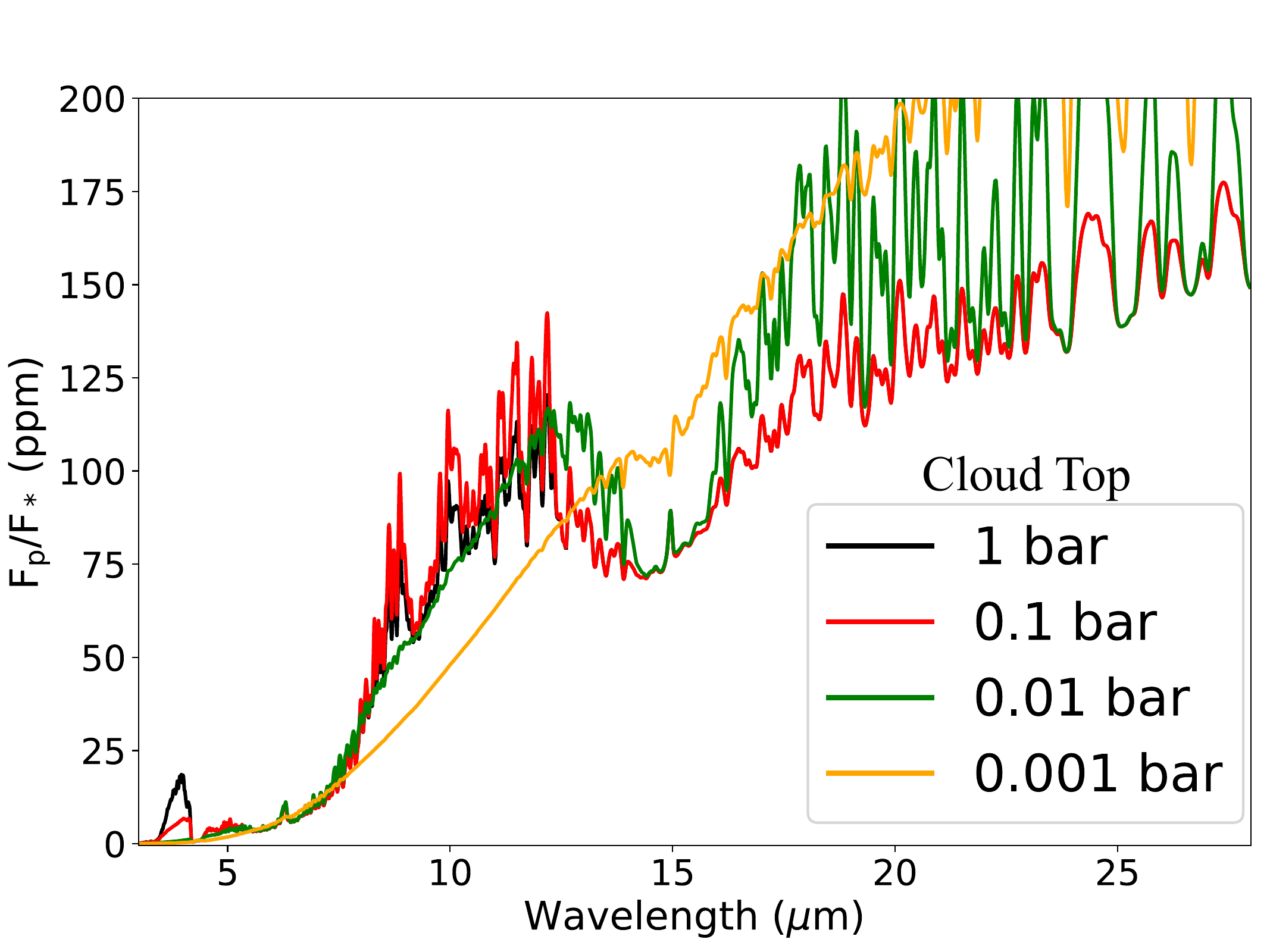}
    \caption{Gray cloud deck  effects on the emission spectra. We color-coded the various pressure levels of the position of the gray cloud layer (see legend in inset).
 }
    \label{fig:cloud_top}
\end{figure}

\begin{figure}[ht!]
    \centering
    \includegraphics[scale=0.43, trim=0cm 0cm 0cm 0cm, clip]{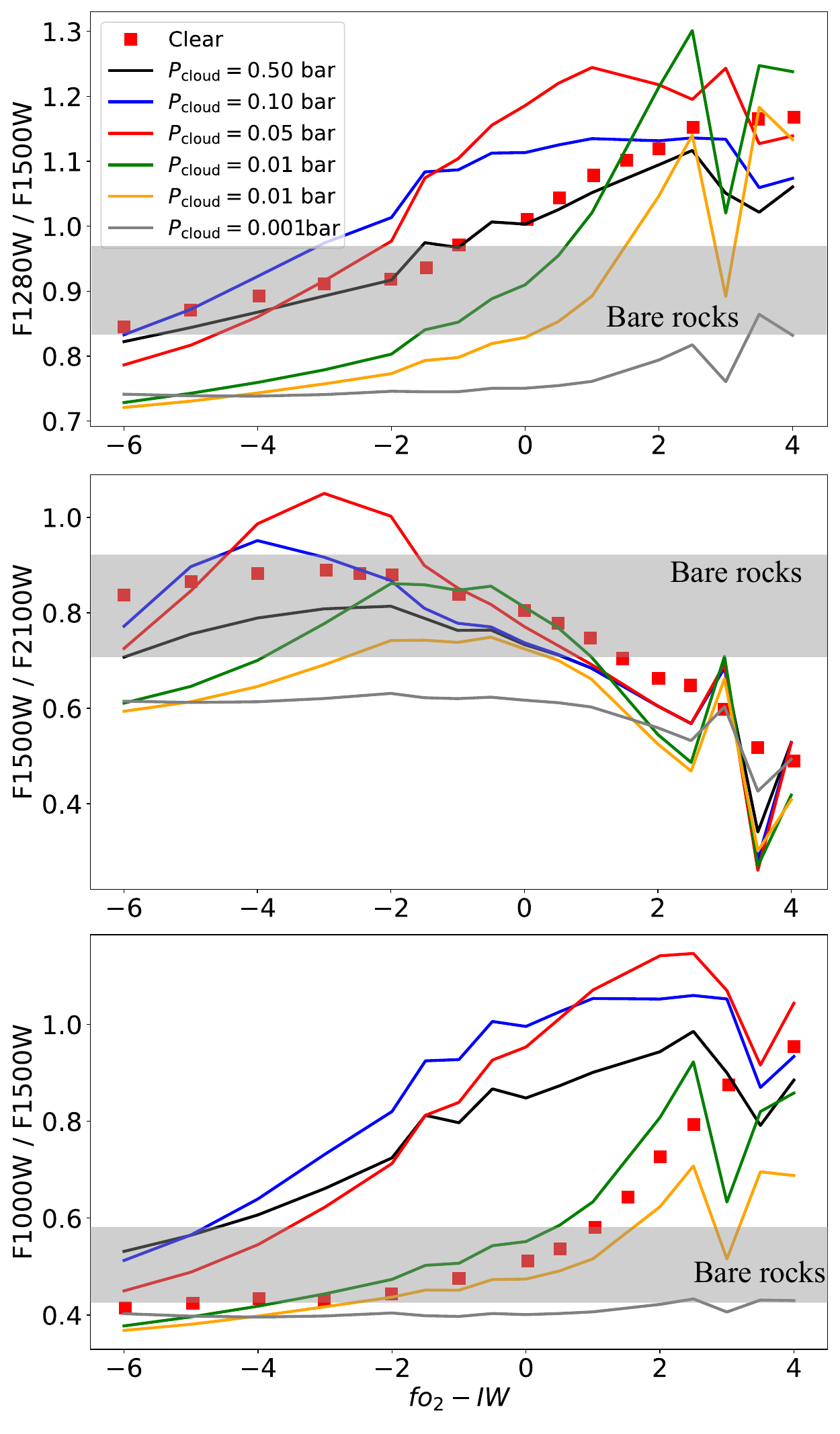}
    \caption{Gray cloud effects in the spectral indexes introduced in Figure \ref{fig:indexes}. With red squares we show the clear atmospheric scenario and with colored lines the gray cloud deck at various pressure levels. With the shadowed gray area, we show the 1$\sigma$ variation of all the bare rock scenarios simulated in this work.  The example shown is the 1 bar surface pressure at various oxidation states for $a_c=10^{-6}$. With this figure we emphasize that gray clouds introduce degeneracies between bare rocks and atmospheres even for highly oxidized atmospheres.
 }
    \label{fig:Index_cloud}
\end{figure}

However, their effect is negligible at higher wavelengths  ($> 5 \, \mathrm{\mu m}$ ).  This indicates that the important factor for our targeted spectral range is $\beta$.  For this reason, we fixed $H_{f} $ in an intermediate value  ($10^3$)  and simulated various scenarios for $\beta$ ranging from -4 to 0.   Where in the limit of 0, the gray clouds are considered at various pressure positions (e.g., 1, 10, 100 mbar). The gray cloud extinction is wavelength-independent.  These decks represent fully optically thick cloud layers that block emission from deeper layers. In this approach, the spectrum is truncated at the cloud-top level, simulating the limiting case of high condensate number density and large optical depth. This is shown in Figure \ref{fig:cloud_top}, where spectral features become increasingly muted as the cloud-top pressure decreases.

These treatments provide a unified and flexible framework to explore aerosol effects, bridging the continuum between photochemical hazes and condensate clouds, as well as from optically thin to thick conditions. However, the formation efficiency, composition, and optical properties of such hazes remain highly uncertain for closed-in tidally locked rocky exoplanets, particularly under the temperature regime of LP 791-18 d. A more detailed treatment is deferred to future studies incorporating coupled photochemistry and microphysics models. 

\begin{figure}[ht!]
    \centering
    \includegraphics[scale=0.25, trim=0cm 2.5cm 0cm 3.5cm, clip]{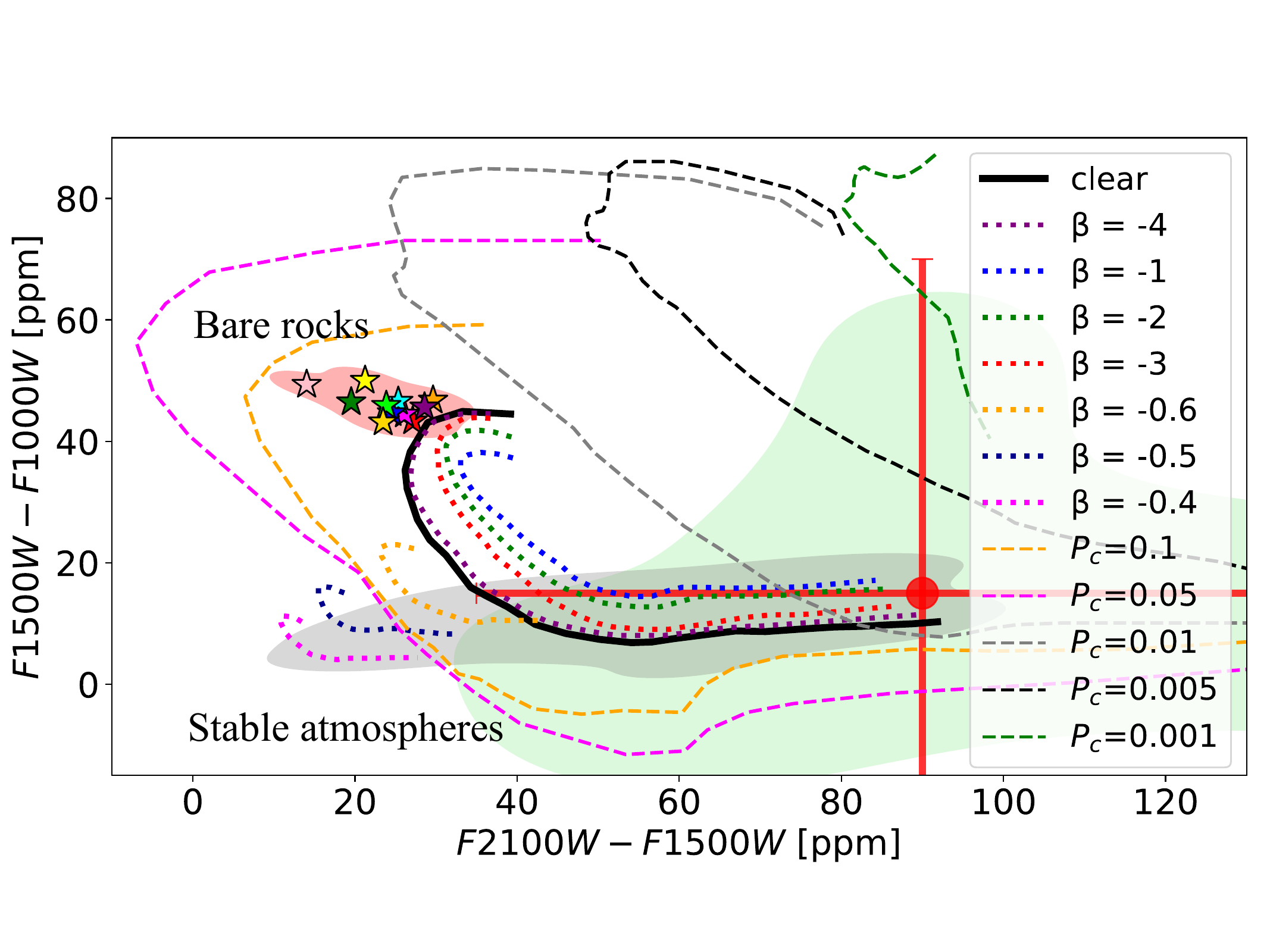}
    \caption{Proposed color-color diagram including the effects of cloud and hazes. The black solid line gives the mean value of the atmospheric scenarios shown in Figure \ref{fig:color} and the colored stars and pink shaded area are the same as in \ref{fig:color}. With dotted colored lines we show the effects of $\beta$ factor from -4 to -0.4 and with gray shadowed region their combined KDE for the stable atmospheres. With dashed colored curves the limit of $\beta \rightarrow 0$ is shown which represents the effect of gray clouds at various pressure layers and the green shaded area is the KDE of the combined gray cloud stable atmospheres scenarios. For reference, a red point with error bars is overplotted, representing the expected observational uncertainty based on propagated errors from the JWST/MIRI secondary eclipse setup proposed by \citet{Benneke2024}. This serves as a benchmark for evaluating the extent to which different scenarios are observationally distinguishable given current instrumental capabilities.
 }
    \label{fig:color_haze}
\end{figure}

In Figure \ref{fig:Index_cloud} we reproduce again the spectral indexes shown in Figure \ref{fig:indexes} only for surface pressure 1 bar as a function of oxygen fugacity and $a_c=10^{-6}$ but now we introduce the effects of gray cloud decks at various vertical layers. Those effects can flatten  the oxygen fugacity trend across the three JWST ratio indices we have experiment  and produce degeneracies in the separation with a bare rock scenarios for the same indices . Finally, for each value of $\beta$ and each gray cloud deck pressure considered, we simulated emission spectra across the entire parameter space, as shown in Figure \ref{fig:color}, and calculated the average trajectory in the color–color space for each scenario affected by photochemical haze or clouds. We present these curves, which encompass the full range of possible effects from haze or clouds, in Figure \ref{fig:color_haze}. The black solid line represents the average of all points from Figure \ref{fig:color}, while the dashed and dotted lines correspond to the average trajectories of simulated points that include cloud or haze effects, respectively. The shadowed regions are calculated in the same manner as Figure \ref{fig:color}, With pink for bare rock clustering, gray the combined population of $\beta$ -4 to -0.4  and green the combined gray cloud population at all pressure levels.  Finally, with a red point we overplot the possible detection, including  error bars, representing the expected observational uncertainty based on propagated errors from the JWST/MIRI secondary eclipse setup proposed by \citet{Benneke2024}. This serves as a benchmark for evaluating the extent to which different scenarios are observationally distinguishable given current instrumental capabilities.

\subsection{Sensitivity of observables to key system parameters} \label{sec:sensitivity}

A central goal of this study was to understand how the physical and chemical properties of LP~791-18\,d influence its observable spectral features, particularly in the context of thermal emission spectroscopy and the photometric capabilities of JWST. Among the various atmospheric characteristics, the most important drivers of emission signatures are the chemical composition and the thermal structure of the atmosphere. The chemical composition, encapsulated by the mean molecular weight, strongly affects atmospheric opacity and therefore the depth and shape of spectral features. In Figure \ref{fig:graphite}, the dependence sensitivity of the mean molecular weight on three key interior-related parameters is shown: oxygen fugacity, graphite activity, and surface pressure. Our analysis shows that oxygen fugacity is the dominant factor, producing large compositional transitions between reduced and oxidized regimes. In contrast, graphite activity and surface pressure introduce subtler variations in molecular weight. However, atmospheric thickness, quantified by surface pressure, emerges as a crucial parameter for spectral observability. As illustrated in Figures \ref{fig:15umVSFO2}  and \ref{fig:indexes}, increasing surface pressure pushes the photosphere to higher altitudes, where temperatures are lower. This results in compressed and muted spectral features, making thick atmospheres appear more spectrally similar to bare-rock scenarios under certain conditions. Lastly, for atmospheric stability and escape, the most critical parameters are the mean molecular weight near the base of the wind and the stellar XUV flux, which together control whether the atmosphere can persist over geological timescales.

The thermal structure is shaped by several factors, but we find that surface albedo plays only a secondary role. As shown in Section~\ref{sec:albedo}, varying the surface albedo mainly affects the lower atmospheric layers, with negligible influence on the temperature profile at higher altitudes, where most of the thermal emission originates. Therefore, surface albedo does not significantly alter the spectral features observed in the mid-infrared.

\section{Conclusions and discussion} \label{sec:discussion}

There is a substantial body of literature on the theory of secondary atmospheres, curated in Earth and planetary sciences over decades \citep[see, e.g.,][]{Gaillard2014}. However, we need more empirical evidence on the early phases of secondary atmosphere generation, along with a diverse sample of the star-planet population. Our current observing capabilities limit our ability to fully characterize the abiotic baseline of a planet, which is crucial for assessing the potential for false-positive biosignature signals. 

An important goal, as community, is to identify selected candidates, such as LP 791-18\,d, that are in a phase similar to early-Earth, where plain outgassing, a process responsible for secondary atmospheres, is amenable to observations from contemporary instrumentation in the upper layers of the atmosphere, where physical processes are out of chemical equilibrium and cooler.  We can make a connection to the interior redox state from those atmospheric observations. In this work, we have shown that the bulk mean molecular weight (\( \mu \)) of the atmosphere is controlled by oxygen fugacity rather than bulk planetary metallicity. Variations in redox conditions determine the dominant molecular species and their partitioning between oxidized and reduced forms, which in turn shape the overall atmospheric composition and $\mu$. We have demonstrated that a variety of atmospheres can be generated from different interior redox states, assuming that the surface pressure of LP 791-18\,d varies from 0.1 to 100 bars. More simulations are needed for similar candidates, in combination with observations, to help disentangle the value of oxygen fugacity. A few promising candidates have been recently identified. For example, \citet{Bello-Arufe2025} present evidence for volcanic activity on the sub-Earth planet L98-59b. A theoretical analysis of a larger sample of similar objects, combined with future observations of atmospheric characterization, could provide insightful information on the planet's chemical inventory relative to stellar abundances. This is a crucial connection when estimating the abiotic baseline of a rocky planet and comparing it with its observable signal to detect potential deviations that might be associated with extraterrestrial life.

\subsection{The bare rock – thick atmosphere degeneracy}
The efforts to characterize small rocky exoplanets have been initiated through JWST observations targeting a handful of such worlds. The observation strategies have primarily focused on secondary eclipse photometry in the \( 15 \, \mathrm{\mu m} \) band, where \( \text{CO}_2 \) shows strong absorption. The rationale behind this observational approach is that comparing the \( 15\, \mathrm{\mu m} \) band with models of bare-rock and \( \text{CO}_2 \)-rich atmosphere scenarios was expected to reveal a distinction in secondary eclipse depths \citep[see, e.g.,][]{Zieba2023,Xue2024}. In the recent work by \citet{Ducrot2024}, they performed observations with one additional photometric MIRI-JWST filter, namely F1280W in the \( 12.8 \, \mathrm{\mu m} \) band, and concluded that a likely bare-rock scenario exists for TRAPPIST-1 b. \citet{Hammond2025} argued that photometric observations in the \( 15 \, \mathrm{\mu m} \) band alone are not sufficient to unambiguously distinguish between a bare-rock planet and an atmosphere, even with the combination of the two MIRI filters F1280W and F1500W. This is because thick atmospheres can still produce thermal emission spectra that resemble nearly featureless blackbodies at these wavelengths. In particular, they significantly suppress the 15$\, \mathrm{\mu m}$ CO$_2$ absorption band. As a result, distinguishing such atmospheres from bare-rock emission becomes challenging with current observational capabilities. Instead, they proposed using F1500W phase curve observations to detect a possible heat redistribution signal to the night side. 

 In this work, we chose LP 791-18\,d as a case study planet and performed comprehensive climate simulations, which were computationally feasible in 1D modeling. These simulations include outgassing at the lower boundary, thermochemistry, and photochemistry coupled with a radiative-convective model to solve for the thermal structure. This was possible because, based on tidal heat flux calculations for LP 791-18\,d, we could constrain its interior melt temperature, which is a critical parameter for our outgassing model. The output of secondary atmospheres was tested for stability, utilizing the criterion of hydrodynamic escape, which was modeled using the isothermal wind approximation. The isothermal approximation was used as a diagnostic tool to explore atmospheric stability across the parameter space defined by surface pressure, oxygen fugacity, and graphite activity.
 Including radiative cooling through atomic and molecular line emission would alter the thermospheric temperature and pressure structure, reducing the efficiency of hydrodynamic escape. Then, the mass-loss rates presented here should be regarded as qualitative upper bounds. This underscores the value of future models incorporating a more detailed treatment of upper atmospheric physics. A more accurate assessment of atmospheric stability could be achieved by coupling our 1D lower atmosphere model with an upper atmosphere module that solves the full energy and momentum equations \citep[see, e.g.,][]{Yelle2004, Tian2005, Johnstone2018}. Nonetheless, the energy-limited approximation employed here already accounts for unresolved processes via the efficiency factor $\eta$, which encapsulates the complexity of upper-atmospheric energy redistribution in a time-independent and agnostic manner \citep{Owen2019}. To avoid overcomplicating the numerical framework, we adopted a simplified approach that still captures the peak temperature of the thermal profile, located near the base of the wind region \citep{Lampon2020}. If the conditions for wind formation are not met, the atmosphere is considered stable. In such cases, Jeans and non-thermal escape processes can shape the chemical inventory of the atmosphere over geological timescales, but they are unlikely to significantly reduce the total atmospheric mass. The sensitivity of atmospheric stability to the planet’s interior redox state is illustrated in several figures. In the bottom panel of Figure \ref{fig:mean_weight}, we show the mean molecular weight at \( R_{\text{xuv}} \), where most of the XUV energy is deposited, and overplot the stable thermosphere solutions. Stability in these models arises from two main factors: the intensity of incoming XUV radiation and the mean molecular weight of the background atmosphere at the wind base. In our case, the XUV energy is constant since we study the current scenarios in time. Atmospheric stability is generally achieved for scenarios with an oxidized redox state in the interior of LP 791-18\,d. This stability occurs for oxygen fugacity values higher than \( \approx 3 \), and for surface pressures approaching \( 100 \) bar, stability is achieved at \( \approx 2 \). For varying graphite content, stability can be achieved starting from oxygen fugacity \( \approx 2 \) for all surface pressures. Our solutions show smaller, stable regions, as seen in Figure \ref{fig:mean_weight}. Although our results highlight these stable regions (as shown in Figure~\ref{fig:mean_weight}), higher-resolution simulations in the redox–pressure parameter space are needed to better delineate their boundaries.
 
For all the simulations that achieved steady-state solutions, we generated synthetic spectra for the secondary eclipse depth (e.g., Figure \ref{fig:emission} for 1 bar mock spectra as a function of interior redox state). In addition, we considered a variety of bare-rock scenarios to cover cases ranging from featureless spectra (e.g., a metal-rich surface) to cases with strong features (e.g., ultramafic). We then examined the effectiveness of various observational strategies. Specifically, for LP 791-18\,d, the F1500W filter is the only one utilized in JWST observations \citep{Benneke2024}. In Figure \ref{fig:15umVSFO2}, we present the secondary eclipse depths as a function of the interior redox state and surface material composition for the \( 15\, \mathrm{\mu m} \) band. In the same figure, we indicate which generated secondary atmospheres can maintain a stable thermosphere, considering the planet’s space environment. Given the estimated age of the LP~791-18 system ($\sim500$ Myr) and our calculated atmospheric mass-loss rates for unstable scenarios ($\sim10^5$--$10^8$~kg\,s$^{-1}$), we find that continuous volcanic outgassing alone is likely insufficient to sustain the atmosphere over geological timescales. To assess this, we estimate the total cumulative atmospheric loss over the planet’s lifetime as follows:

\begin{equation}
\begin{split}
\Delta M = \dot{M} \times t &\approx (10^5\,\mathrm{to}\,10^8~\mathrm{kg\,s}^{-1}) \\
                            &\times (500 \times 10^6~\mathrm{yr}) \approx (1.6 \times 10^{21}~\mathrm{to}~1.6 \times 10^{24})~\mathrm{kg}.
\end{split}
\end{equation}

At the lower end of this range, the total mass lost approaches, or even exceeds, typical estimates for the initial volatile inventory of rocky planets. The initial volatile budget of terrestrial planets is poorly constrained but is generally estimated to be in the range of $10^{-5}$ to $10^{-2}$ of the planetary mass \citep{Elkins-Tanton2008,Zahnle2010,Schaefer2010}. For Earth, this corresponds to a bulk silicate inventory of approximately $(1.8$--$3.0) \times 10^{22}$~kg of volatiles, primarily in the form of H$_2$O, CO$_2$, and other light elements \citep{Hirschmann2006,Dasgupta2010}. For LP~791-18\,d, a conservative volatile reservoir on the order of $10^{-3} M_p$ would yield $\sim 6 \times 10^{21}$~kg of volatiles. This estimate suggests that atmospheric escape at mass-loss rates of $10^7$--$10^8$~kg\,s$^{-1}$ would exhaust such a reservoir in less than $\sim$200\,Myr. Even for lower escape rates ($10^5$--$10^6$~kg\,s$^{-1}$), the volatile budget would be significantly eroded within 1--2\,Gyr. Furthermore, the XUV activity of the host star is expected to be substantially higher in its early evolutionary stages. Therefore, the long-term persistence of an atmosphere on LP~791-18\,d requires either (1) formation with an unusually large volatile inventory (e.g., from a carbon-rich or water-rich disk), or (2) a redox state that favors the buildup of a dense, high-mean-molecular-weight atmosphere capable of resisting hydrodynamic escape. Under these assumptions, we conclude that reduced interior scenarios (low $f_{\mathrm{O}_2}$) are unlikely to retain secondary atmospheres over gigayear timescales, especially if the outgassed composition is dominated by light molecules. This reinforces our finding that only oxidized atmospheric compositions (with $f_{\mathrm{O}_2}-\mathrm{IW} \gtrsim 2$) produce atmospheres that are stable against hydrodynamic loss.

Nevertheless, our stability criterion is based on approximations. While we cannot determine with precision the exact redox threshold below which atmospheres undergo continuous hydrodynamic escape throughout their lifetimes, our results suggest that reduced atmospheres with \( f_{\mathrm{O}_2} - \mathrm{IW} \lesssim -1 \) are likely to be dominated by very light species, both in total atmospheric mass and at \( R_{\text{xuv}} \). In the absence of other mitigating mechanisms, such atmospheres are unlikely to suppress the planetary wind regime, except through a significant decline in the stellar XUV flux over time. For redox states between \( 0 \lesssim fO_2 - IW \lesssim 2 \), even though our current framework predicts planetary wind regimes, the atmospheric composition is rich in heavier elements like carbon and oxygen. Thus, due to molecular diffusion in the upper atmosphere, which preferentially allows light elements to escape, we expect the bulk atmospheric composition to become progressively enriched in heavier species over geological timescales. This could potentially stop the hydrodynamic escape flow until degassing from the lower boundary brings the atmospheric composition to a state where escape is initiated once more. In the absence of hydrodynamic escape, the remaining mass loss mechanisms, Jeans escape and non-thermal processes, can still cause significant compositional evolution across Gyr timescales. These processes preferentially remove lighter species (e.g., H$_2$, He, CH$_4$), enriching the upper atmosphere in heavier molecules and atoms (e.g., CO$_2$, O$_2$, N$_2$). This selective loss increases the mean molecular weight near the exobase, which in turn reduces escape efficiency and may help stabilize the atmosphere over time (\citet{Zahnle2017}). As a result, the present-day atmospheric composition may differ from the initially outgassed one, especially in intermediate redox states scenarios, where both light and heavy species are outgassed in the atmosphere. This evolutionary effect should be considered when interpreting spectral features and mean molecular weight as indirect diagnostics of the planet’s interior redox state.

For all the above mock spectra, we utilized PANDEXO \citep{Batalha2017} to estimate the observational uncertainties, given the characteristics described in \citet{Benneke2024}. The uncertainties are estimated to be on the order of 30-45 ppm, depending on the scenario. From Figure \ref{fig:15umVSFO2}, we can expect to distinguish an atmospheric detection if the atmosphere is highly oxidized and thin or highly oxidized with low graphite content \( a_c \lesssim 10^{-4} \). Nevertheless, it would be difficult to make a bold claim even for a thin atmosphere (\( \approx 0.1 \) bar) given the estimated uncertainties. Following the work of \citet{Ducrot2024}, we investigated the combination of two photometric measurements from the available MIRI filter list. After performing numerical experiments, we show the best combinations in Figure \ref{fig:indexes}. We display the ratios of the filters as a function of the interior redox state. For the combination of filters F1280W-F1500W, thin atmospheres show differences of up to \( 25\% \) in secondary eclipse depth, which is at the limit of detection, considering the uncertainties and the thermal contrast estimated for LP 791-18\,d.  For thick atmospheres, the ratio approaches unity, confirming the degeneracy observed in \citet{Hammond2025}. If we consider alternative pairs of filters, F2100W-F1500W and F1000W-F1500W, we obtain similar results to the first pair. However, we can achieve better contrast in the filter fraction and can more effectively separate at least the thin atmospheres. For thick atmospheres, the signal is suppressed. Additionally, degeneracies can be produced even for thin atmospheres if we add effects of thick photochemical haze or gray cloud deck at low pressure levels as it is shown in Figure \ref{fig:Index_cloud}.  Our simulations show that we cannot  unambiguously distinguish between an atmosphere and bare rock with observations of one or even two photometric bands.

Finally, we propose a novel approach, presented in Figure \ref{fig:color}, which utilizes three MIRI photometric filter observations and plots the estimated results as a color-color diagram. In this diagram, all the bare-rock scenarios cluster in the upper-left corner, while the stable, highly oxidized atmospheres are located in the bottom-right corner. Surprisingly, when comparing these three photometric bands, the dependence on surface pressure and graphite activity is eliminated, since most of the plotted points follow a smooth slope with small dispersion. Due to the non-linearity of the redox state and the strength of individual spectral bands, the intermediate region of the color-color points is mixed with both reduced and oxidized scenarios, while the extremes are well-separated, as expected. Similarly, the clustering of reduced atmospheres is observed, as they do not exhibit strong spectral features in contrast to oxidized atmospheres, as shown in Figure \ref{fig:emission}. A real observation would likely result in a measurement that lies in one of the two shadowed areas. With the estimated uncertainties of \( \pm 30-45 \, \text{ppm} \), it is expected to be able to separate an atmosphere from an airless case for LP 791-18\,d, regardless of its surface pressure or graphite content. We performed a parametric study of photochemical hazes and both wavelength-dependent semi-gray and gray clouds across all simulations presented in this work and examined their effects on our proposed color–color diagram, shown in Figure \ref{fig:color_haze}. Our results show that, while variations in the properties of photochemical hazes, semi-gray, and gray clouds can modify the trends followed by different atmospheric scenarios in the diagram, the region corresponding to highly oxidized, stable atmospheres remains clearly distinguishable from that of bare rock surfaces. Based on the uncertainty estimation and its propagation for constructing the color–color diagram, we overplotted a representative synthetic detection, as shown in Figure~\ref{fig:color_haze}. In order for the proposed color–color system to be observationally useful in the JWST era, the typical uncertainty required is on the order of $\sim 50$ ppm, which is currently achievable only for highly oxidized atmospheres.Finally, we emphasize that the proposed framework provides a foundation for prioritizing future observations, particularly as instrumental capabilities and retrieval techniques advance.

\subsection{The importance of stable thermospheres in habitability and detectability of rocky exoplanets.  } \label{thermosphere}
As we discussed above, stable thermospheres retain the atmospheric mass against bulk escape and play a crucial role in the habitability of rocky planets. This stability primarily results from the combination of two physical parameters: stellar incoming radiation and the mean molecular weight. While these parameters help stabilize the atmosphere, they pose challenges for detectability. Due to cooler atmospheres with high mean molecular weight, detecting such atmospheres becomes extremely challenging, even for future-generation detectors. In transmission, as mentioned above, the spectral feature is proportional to the atmospheric scale height parameter, given by $\delta D \approx \frac{2 R_p H}{R_*^2}$ where \( H \) depends on the temperature and \( R_p \) is the radius where the optical depth of the spectral wavelength is \( \approx 1 \). It is thus evident that alternative approaches must be considered that can enhance the detectability of small rocky planets. 

When comparing the photoabsorption cross-sections of various species across different spectral regions, UV absorption (particularly in the FUV–NUV range) is typically many orders of magnitude stronger than in the optical to near-infrared \citep[see, e.g.,][]{Heays2017,Venot2018,Malik2019}. The higher-energy photons of the stellar spectrum are absorbed in the upper atmosphere (e.g., from the thermosphere to the exosphere), such as the \( R_{\text{xuv}} \) concept used in Section \ref{sec:stability}.  Therefore, it is evident that transit radius in the FUV-NUV-XUV wavelengths are optimizing our probabilities for characterization  of small rocky exoplanets with transmission spectroscopy. 

Furthermore, stable upper atmospheres have one more characteristic that makes them appealing targets for future observations: their thermal structure is easier to model than the lower atmosphere. This is because the dominant factors in the energy balance equation are the incoming UV energy as the heating mechanism and thermal conduction as the cooling mechanism \citep{Bougher1991,Gkouvelis2021}. The thermospheric thermal structure has the morphology of an inversion profile, which leads to an isothermal layer at the exosphere \citep{Banks1973,Brecht2022}.  Typical temperatures in this region range from a few hundred to a few thousand kelvins, which further adds to their observability. These components benefit future UV transmission spectroscopic observations when coupled with Bayesian inference frameworks to characterize small rocky exoplanets. While no detector is nowadays capable of performing these observations, an effort is ongoing to include a UV spectropolarimeter on the future Habitable Worlds Observatory mission \citep[see,][]{Muslimov2024}.

\begin{acknowledgements}

 L.G. acknowledges the "Severo Ochoa excellence" visitor program for partially funding this study. Author F.J.P acknowledges financial support from the Severo Ochoa grant CEX2021-001131-S MICIU/AEI/10.13039/501100011033 and Ministerio de Ciencia e Innovación through the project PID2022-137241NB-C43

\end{acknowledgements}

\bibliographystyle{aa} 
\bibliography{aa54192-25} 

\end{document}